\documentclass [a4paper,11pt]{article}
\usepackage[english]{babel}
\usepackage{graphicx,amsmath,amsfonts,amssymb,slashed,upgreek,color,caption,amscd,cite,cancel,footnote}
\usepackage{multirow}
\usepackage{graphicx,subfigure}
\DeclareGraphicsExtensions{.pdf,.png,.jpg}

\renewcommand{\arraystretch}{1.3}

\newcommand{\be}{\begin{equation}}
\newcommand{\ee}{\end{equation}}
\newcommand{\beq}{\begin{equation}}
\newcommand{\eeq}{\end{equation}}
\newcommand{\bea}{\begin{eqnarray}}
\newcommand{\eea}{\end{eqnarray}}

\newcommand{\R}{{}^{(3)}\!R}

\newcommand{\om}{\Omega}

\newcommand{\Oee}{\Omega_{m}}

\newcommand{\we}{\bar w}
\newcommand{\Oe}{\bar \Omega_m}
\newcommand{\x}{x}
\newcommand{\xo}{x_{0}}
\newcommand{\aaa}{\alpha}
\newcommand{\bb}{\beta}
\newcommand{\Mb}{M}
\newcommand{\Lb}{\lambda}
\newcommand{\cb}{{\cal C}}

\usepackage[stable]{footmisc}

\def\MM{M_{*}}

\begin{document}

\begin{center}
\Large{\textbf{Phenomenology of dark energy: \\
exploring the space of theories with future redshift surveys}} \\[1cm]
\large{Federico Piazza$^{\rm a}$, Heinrich Steigerwald$^{\rm b}$ and Christian Marinoni$^{\rm b,c}$ }
\\[0.5cm]

\small{
\textit{$^{\rm a}$  APC, (CNRS-Universit\'e Paris 7), 10 rue Alice Domon et L\'eonie Duquet, 75205 Paris, France \\ 
}}

\small{
\textit{ PCCP, 10 rue Alice Domon et L\'eonie Duquet, 75205 Paris, France \\ 
}}

\vspace{.2cm}

\small{
\textit{$^{\rm b}$ Aix Marseille Universit\'e, CNRS, CPT, UMR 7332, 13288 Marseille,  France. \\
}}

\small{
\textit{Universit\'e de Toulon, CNRS, CPT, UMR 7332, 83957 La Garde,  France. \\
}}

\vspace{.2cm}

\small{
\textit{$^{\rm b}$ Institut Universitaire de France, 103, bd. Saint-Michel, F-75005 Paris, France
}}

\end{center}

\vspace{2cm}

\begin{abstract}

We use the effective field theory of dark energy to explore the space of modified gravity models which are capable of driving the present cosmic acceleration. 
We identify five universal functions of cosmic time that are enough to 
describe a wide range of theories containing a single scalar degree of freedom in addition to the metric. The first function (the effective equation of state) uniquely controls the expansion history of the universe. The remaining four functions appear in the linear cosmological perturbation equations, but only three of them regulate the growth history of large scale structures. We propose a specific 
parameterization of such functions in terms of characteristic coefficients that serve as coordinates in the space of modified gravity theories and can be effectively constrained by the next generation of cosmological experiments. 
We address  in full generality the problem of the soundness of the theory  against ghost-like and gradient instabilities and show how the space of non-pathological models shrinks when 
a more negative equation of state parameter is considered.  This analysis allows us to locate  a large class of  stable theories  that violate  the null energy condition (i.e. super-acceleration models) and to 
recover, as particular subsets, various  models considered so far.  
Finally, under the assumption that the true underlying cosmological model is the $\Lambda$ Cold Dark Matter ($\Lambda$CDM) scenario, and relying on the figure of merit of EUCLID-like observations, we demonstrate that the theoretical requirement of stability significantly narrows the empirical likelihood, increasing the discriminatory power of data. We also find that the vast majority of these non-pathological theories generating the same expansion history as the $\Lambda$CDM model predict a different, lower, growth rate of cosmic structures. 
\end{abstract}

\newpage 
\tableofcontents

\vspace{.5cm}


\section{Introduction}
Understanding the present accelerating phase of the universe in terms of fundamental physics is an outstanding challenge of modern cosmology.
Alternatives to the $\Lambda$CDM paradigm, the so called dark energy models,   date at least as early as the discovery~\cite{Perlmutter:1998np, Riess:1998cb} of the acceleration itself. In the simplest \emph{quintessence} models~\cite{wett:1988, Sahni:1999gb, Caldwell:1997ii, Binetruy:2000mh, Peebles:2002gy}, the required negative pressure is produced by a scalar field rolling down its potential.  Dark energy models well beyond minimally-coupled quintessence have also  been  explored. Proposals range from coupled quintessence~\cite{Brans:1961sx,Amendola:1999er,Uzan:1999ch, Perrotta:1999am, Riazuelo:2001mg,Gasperini:2001pc,Perivolaropoulos:2003we, pettorino} to extra-dimensional  mechanisms~\cite{DGP, Deffayet:2001pu}, from considering higher curvature terms in the Lagrangian (such as in $F(R)$~\cite{Starobinsky:1980te,DeFelice:2010aj, Sotiriou:2008rp,Amendola:2007,Capozziello:2003tk} or $F(G)$~\cite{Nojiri:2005jg} theories), as well as torsion terms~\cite{Li:2011}, to models of non-local~\cite{Arkani-Hamed:2002fu,Dvali:2007kt,Deser:2007jk}  and massive~\cite{deRham:2010kj,Hassan:2011hr,Comelli:2013tja} gravity or departures from the geometrical description of general relativity~\cite{Piazza:2009bp,Piazza:2012xh} (see \cite{cope,lucashin,costas} for reviews). 

In the presence of a modification of gravity, 
effects of dark energy are expected not only at the background level---for instance, on the Hubble rate $H(t)$---but also at the level of  cosmological perturbations---for instance, in the
growth rate  $f(t)$  of large scale structures.\footnote{The linear growth rate function is  $f(t)=\frac{d \ln \delta}{d \ln a(t)}$, where $\delta$ represents the fractional overdensity of matter}
Current observational programs, which constrain $f(t)$ with a $15\%$ precision in the redshift range up to $z \sim 1.3$  \cite{Guzzo2008, SonPer09, DavNusMas11, BlaGlaDav11, ReiSamWhi12, SamPerRac12, BeuBlaCol12, TurHudFel12, Torre_2013}, already provide interesting evidence for ruling out the most extreme proposals of modified gravity. Future surveys such as DES\cite{des}, LSST\cite{lsst}, BigBoss~\cite{bigboss}, EUCLID\cite{euclid1,euclid2}  are thus looked to with 
expectant attention, as they will eventually attain the necessary precision to challenge also the finest deviations from standard gravity predictions on large cosmological scales. 

However, empirical precision is not the only fundamental goal that a measuring protocol must satisfy in cosmology \cite{Peebles:2002iq}.
Since measurements are indeed estimations of parameters using a theory, testing the soundness of the theoretical framework that links
physical  observables to cosmological parameters is also of critical importance. For example, 
direct and independent measurements of the dark energy equation of state parameter~\cite{astier, marpairs, mod1, Bel:2012ya, san,anderson, Ade:2013zuv} on the one hand and of $f(t)$ on the other                       
~\cite{Guzzo2008, SonPer09, DavNusMas11, BlaGlaDav11, ReiSamWhi12, SamPerRac12, BeuBlaCol12, TurHudFel12, Torre_2013}  inevitably loose track of 
the specific mechanisms responsible for the possible deviations from general relativity and, ultimately, of the underlying theory. The orthogonal strategy---\emph{i.e.} assessing, on an individual basis,  
the observational viability of specific, more or less physically motivated models---is far from economical, and prevents to make model-independent statements about unknown regions of theory-space.

An intermediate  perspective is recommendable: developing a consistent formalism that can incorporate all the possible gravitational laws generated by 
adding a single scalar degree of freedom to Einstein's equations. 
Such a phenomenological approach should efficiently keep track of both the background behavior (\emph{i.e.} $H(t)$) and the dynamics on smaller scales (\emph{i.e.} the cosmological perturbations) responsible, for example, for the growth rate $f(t)$. Several strategies have been proposed along this direction. In this paper we will make use of the \emph{effective field theory} (EFT) \emph{of Dark Energy} proposed in~\cite{GPV} and further developed in~\cite{BFPW,GLPV,Bloomfield:2013efa}, which extends to late-time cosmology the formalism of the EFT of inflation~\cite{EFT1,EFT2} (see also~\cite{Creminelli:2008wc} for the treatment of a more restricted class of models and~\cite{PV} for a review). Other notable strategies include the parameterized post-Friedmaniann formalism~\cite{Baker:2011jy,BFS,Baker:2013hia}, the covariant EFT approach in its various versions~\cite{PZW,BF,Battye:2012eu}, the imperfect fluid approach~\cite{Sawicki:2012re}.

Instead of parametrizing a specific theory in terms of variables, letting to observations the task of fixing their amplitudes, 
the EFT of dark energy makes it possible to parametrize  theories themselves in terms of \emph{structural functions} of time. The advantage is that  one can interpret observations directly in the phase space of theories and not within the framework of a single paradigm. The price to pay is that,  instead of fixing numbers,  observations should have enough power to fix continuous functions of time.
The apparent intractability of the problem
can be finessed by  phenomenological modeling, i.e. by  compressing  the information contained in the structural  functions into a finite set of coefficients. The tricky step is to engineer 
a  parametrization  which can be  effectively constrained by observations, yet it is   flexible and universal enough to allow exploring most of the phase  space of stable theories, i.e. models that 
do not suffer from ghosts instabilities.   

In this paper we present a specific way to address this challenge and show how the EFT of dark energy can efficiently confront the observations. 
The goal is to show what constraints on theoretical scenarios of modified gravity a precise measurement of $H(t)$ and  $f(t)$   can provide.
To this purpose we exploit the growth index formalism  developed in \cite{SBM} which  provides a flexible parameterization of the 
linear growth rate $f(t)$ and a straightforward  mapping of observational constraints from the space of cosmological observables into the phase space  of all possible theories. 

The paper is organized as follows. In Sec.~\ref{sec:2} we review the EFT of dark energy. Our starting point is an action that depends on six structural functions of the cosmic time, and general enough to contain all scalar tensor theories with equations of motion up to second order in derivatives. We quote the background evolution equations as well as the expressions of characteristic  linear perturbation theory observable: the  effective Newton constant $G_{\rm eff}$ and the gravitational slip $\gamma_{\rm sl}$ parameter.  We also discuss what constraints  theoretical consistency, as well as  gravitational tests on various scales,
put on  these structural  functions.    Only three of them appear both in the equations of motion for the \emph{background} and \emph{perturbation}  sectors of the theory, the remaining uniquely characterizing  the \emph{perturbation} equations. However, a residual degeneracy affects the formalism, in that different combinations of the three  functions governing the background can reproduce identical expansion histories. In Sec.~\ref{obvia} we show how we  tackle this issue: by a simple ``back-engineering" procedure, we re-express the three background functions in terms of an \emph{effective equation of state} parameter $\we$~\cite{post-fried1,post-fried2}, which rules the expansion history,  and a coupling function that only governs the perturbation sector. 
Despite the  EFT  formalism  is now reduced to five functions of time, in order to confront observations, we need to implement a parameterization scheme.  In Sec.~\ref{sec:4} we propose a specific parameterization that is by no means unique, but  present a sufficient degree of  generality and   phenomenological merits.
Our formalism allows a neat treatment of the stability issue for dark energy theories (Sec.~\ref{sec:4.3}) and allows to address in full generality the issue of super-acceleration within a consistent, non-pathological  theory. 
Finally, in Sec.~\ref{sec:5} we present the constraints that current measurements of a  particular observable of the perturbation sector, the growth rate of cosmic structures, 
put on the phase space of gravitational   theories and forecast those that a future mission such as EUCLID will provide. Conclusions are drawn in Sec. 6.

\section{The effective field theory of dark energy} \label{sec:2}

The formalism at the basis of this paper (see~\cite{PV} for a review) was first used in Ref.~\cite{ghost} and then applied to Inflation in Refs.~\cite{EFT1,EFT2}. The idea at the basis of the ``effective field theory of Inflation" is to consider cosmological solutions as states with spontaneously broken time-translations, and cosmological perturbations as the associated Nambu-Goldstone excitations. This allows a systematic and unambiguous expansion of the inflationary Lagrangian in operators containing an increasing number of cosmological perturbations. The formalism was then extended to quintessence~\cite{Creminelli:2008wc} and to the most general class of single scalar field dark energy models in~\cite{GPV} (see also~\cite{BFPW}). Later relevant developments include~\cite{GLPV,Bloomfield:2013efa}.

\subsection{The EFT action}
Our starting point here is the following action\footnote{With a slight change of notation with respect to the above cited works on the EFT of dark energy, here we define our structural functions by pulling out the bare Planck mass squared from the Lagrangian. It is straightforward to compile a dictionary between our notation and that of, \emph{e.g.}, Ref.~\cite{GLPV}:    
\begin{subequations} \label{dictionary}
\begin{align}
\Mb^2(t) &= \MM^2 f(t)\, , \qquad \qquad \Lb = \frac{\Lambda}{\MM^2 f(t)}\, ,  \ \qquad \qquad \cb = \frac{c}{\MM^2 f(t)}\, , &\\[2mm]
\mu^2_2 &= \frac{M_2^4}{\MM^2 f(t)}\, ,  \ \quad \qquad \mu_3 = \frac{m_3^3}{\MM^2 f(t)}\, , \, \qquad \qquad \epsilon_4 =  \frac{2 m_4^2}{\MM^2 f(t)}\, .&
\end{align}
\end{subequations}
The scale of the new coefficients is set by appropriate powers of the Hubble parameter
(\emph{e.g.} $\cb\sim H^2$, $\mu_3 \sim H$, $\epsilon_4 \sim 1$). Note also that, in the notations of~\cite{GLPV}, we have set $m_4^2 = \tilde m_4^2$.}

\be \label{example}
\begin{split}
S \ =\  & \ S_m[g_{\mu \nu}, \Psi_i] \ +\ \int \! d^4x \, \sqrt{-g} \, \frac{\Mb^2(t)}{2} \, \Big[R \, -\,  2 \Lb(t) \, - \, 2 \cb(t) g^{00}   \Big. \\[1.2mm]
& \left.+ \, \mu_2^2(t) (\delta g^{00})^2\, -\, \mu_3(t) \, \delta K \delta g^{00} 
   + \,  \epsilon_4(t) \left(\delta K^\mu_{ \ \nu} \, \delta K^\nu_{ \ \mu} -  \delta K^2  +  \frac{\R\,   \delta g^{00}}{2} \right) 
  \right] \;. \\
\end{split}
\ee
Those who are not already familiar with the formalism may not find the above expression particularly illuminating. While referring to the already cited literature for more details, we summarize few main features below. The pragmatic reader may also skip directly to the formulas relating the structural functions $\Mb^2$, $\Lb$, $\cb$, $\mu^2_2$, $\mu_3$, $\epsilon_4$ to the 
background  and perturbation  quantities, in Secs.~\ref{sec_background} and~\ref{perturbedd} respectively.

\begin{enumerate}
\item
The action is specifically tailored for cosmology and written directly in \emph{unitary gauge}: the time coordinate $t$ is fixed to be proportional to the scalar field, while the three space coordinates $x^i$ are left undetermined. This explains the presence of non covariant terms such as the perturbations of the \emph{lapse} component of the metric, $\delta g^{00} \equiv g^{00} + 1$, the perturbation of the  extrinsic curvature on the $t = const.$ hypersurfaces $\delta K_{\mu \nu}$ and its trace, $\delta K$, the three dimensional Ricci scalar $\R$ calculated on such an hypersurface. This choice of gauge also explains the ``disappearance" of the scalar field: its dynamics is entirely encoded in the metric's degrees of freedom.
\item 
The displayed operators reproduce~\cite{GLPV} the entire class of Horndeski~\cite{horndeski} or \emph{generalized Ga\-li\-le\-on}~\cite{Deffayet:2009mn} theories, which are the most general scalar tensor theories not giving rise to derivatives beyond second order in the equations of motion. A large class of dark energy models can be recast in this form (see Table~\ref{table1}).
The structural functions $\Mb^2$, $\Lb$, $\cb$, $\mu^2_2$, $\mu_3$, $\epsilon_4$ are universal, in the sense that they are unaffected by field redefinitions. As shown in~\cite{GPV,GLPV}, action~\eqref{example}  can always be recast into  covariant form.
\item
Violations of the weak equivalence principle are assumed to be negligible or at least irrelevant for the problem at hand. Thus, the action is written in the \emph{Jordan Frame}, \emph{i.e.} in terms of the metric to which matter fields (baryons and dark matter, contained inside $S_m$) are minimally coupled. This is also the metric of most direct physical interpretation~\cite{NP}. 
\end{enumerate}

\begin{table}
\centering
  \begin{tabular}{|p{4.5cm}||c|c|c|c|c|c|}
 \hline
 & & &&&&\\[-4mm]
    &
    $\mu = \frac{d \log \Mb^2(t)}{dt}$ &
    $\ \ \Lb$ \ \ &
    $\ \ \ \cb \ \ $ \ &
    $\ \ \ \mu_2^2\ \ $\  &
    $\ \ \ \mu_3\ \ $\  &
    $\ \ \ \epsilon_4 \ $ \  
    \\[1mm] \hline
       \hline
       $\Lambda$CDM & 0 & const. & 0 & 0 & 0 & 0   \\ \hline
       Quintessence & 0 & \checkmark & \checkmark & 0 & 0 & 0   \\ \hline
       $k$-essence~\cite{ArmendarizPicon:2000dh} & 0 & \checkmark & \checkmark & \checkmark & 0 & 0   \\ \hline
       Brans-Dicke~\cite{BD,polarski} & \checkmark & \checkmark & \checkmark & 0 & 0 & 0   \\ \hline
       $f(R)$ ~\cite{DeFelice:2010aj} & \checkmark & \checkmark & 0 & 0 & 0 & 0   \\ \hline
       Kinetic  braiding~\cite{deffa1}& 0 & \checkmark & \checkmark & \checkmark & \checkmark & 0   \\ \hline
       DGP~\cite{DGP} & \checkmark & \checkmark & \checkmark & \checkmark & \checkmark & 0   \\ \hline
       Galileon Cosmology~\cite{justin} & \checkmark & \checkmark & \checkmark & \checkmark & \checkmark & 0   \\ \hline
       $f(G)$ -Gauss-Bonnet~\cite{Nojiri:2005jg}& \checkmark & \checkmark & \checkmark & \checkmark & \checkmark & \checkmark   \\ \hline
       Galileons~\cite{NRT,Deffayet:2009wt} & \checkmark & \checkmark & \checkmark & \checkmark & \checkmark & \checkmark    \\ \hline
       Horndeski~\cite{horndeski,Deffayet:2009mn}& \checkmark & \checkmark & \checkmark & \checkmark & \checkmark & \checkmark    \\ \hline
  \end{tabular}
  \normalsize
  \caption{
   By expanding on Table 4 of Ref.~\cite{BFPW}, we simply list (see Refs.~\cite{GPV,BFPW,GLPV,Bloomfield:2013efa} for details) some of the explicit dark energy models that are covered by action~\eqref{example}. There are cases such as DGP, $f(G)$ and Galileons in which a specific relation between the listed coefficients are implied.}
\label{table1}
\end{table}

The main advantage of the above gauge choice is 
a neat separation between the terms contributing to the background evolution and those affecting only the perturbations. All terms in the second line of~\eqref{example} are quadratic in the perturbations and hence do not interfere with the background evolution.  The latter is determined uniquely by the three time-dependent functions $\Mb^2(t)$, $\cb(t)$ and $\Lb(t)$. This is a general result that is demonstrated to hold~\cite{EFT2,GPV} for arbitrarily complicated covariant dark energy Lagrangians, as long as they contain only up to one additional scalar degree of freedom. 
The relevant equations for the evolution of background and perturbed cosmological observables are summarized in the next subsections.

\subsubsection{Background sector}\label{sec_background}

In the EFT formalism the background evolution is  governed only by the three functions $\Mb^2(t)$, $\cb(t)$ and $\Lb(t)$ appearing in the first line of~\eqref{example}. This applies to all dark energy theories---no matter how complicated---with up to one scalar degree of freedom. This non-trivial result has been proved in~\cite{GPV}. 
Since the matter fields are essentially constituted by non-relativistic species, we adopt  the perfect fluid approximation and set $p_m \simeq 0$. In a flat universe, the background Einstein equations derived from~\eqref{example} read
\begin{align}
\cb \ &  =   \   \frac12 ( H \mu - \dot \mu - \mu^2 )  + \frac{1}{2 \Mb^2} (\rho_{D}+p_{D}) \;, \label{c2}\\
\Lb \, &  = \ \frac12 (5 H \mu + \dot \mu + \mu^2  ) + \frac{1}{2 \Mb^2}  (\rho_{D}- p_{D})     \;. \label{L2}
\end{align}
where $H= \dot a(t)/a(t)$ is the Hubble expansion rate and we have defined the non-minimal coupling function\footnote{Our coupling $\mu$ corresponds to  $\dot f/f$ in the notations of~\cite{GPV,GLPV}.}
\begin{equation}\label{mudef}
\mu \ \equiv \ \frac{d \log \Mb^2(t)}{dt} \, .
\end{equation} 
The dark energy  density $\rho_D$ and pressure   $p_D$ are  \emph{defined} by the relations  
\begin{align}
 H^2  \  &= \  \frac1{3 \Mb^2(t)} (\rho_m + \rho_{D}  )  \label{frie1}\; ,\\
\dot H  \ &=  \ - \frac1{2 \Mb^2(t)} (\rho_m + \rho_{D} + p_{D}  )  \label{frie2}\;.
\end{align}
Since we are working in the Jordan frame, non-relativistic matter always scales as  $\rho_m \propto a^{-3}$ by energy-momentum conservation.  On the opposite, because of the coupling to gravity,  the energy momentum tensor of dark energy   is not univocally defined. From the above relations we can derive the modified conservation equation       
\begin{equation} \label{conservation}
\dot \rho_D + 3 H (\rho_D + p_D) = 3 \mu \Mb^2 H^2  \, .
\end{equation}

From the above set of equations it is appearent that in the limit $\mu =0$ ($\Mb^2$  constant), the background evolution (\emph{i.e.} $H(t)$) completely determines the remaining structural functions $\cb(t)$ and $\Lb(t)$. Such a limit was specifically considered in Ref.~\cite{Creminelli:2008wc}. However, in general, we need one more input in order to completely determine the background sector. For instance, we can define the equation of state of dark energy $w$ as
\begin{equation}\label{om0}
p_D (t)\, =\,  w(t) \rho_D (t)\, . \qquad 
\end{equation}
Then, if the functions $w(t)$ and $\mu(t)$ are known,  a measure of the constant $H_0$ is enough to close the system and determine the values of $\cb$ and $\Lb$ univocally.

While the background functions completely determine the expansion history $H(t)$, the converse is not true. Indeed, from Eq.~\eqref{frie1} we see that different choices of $\Mb(t)$ and $\rho_D(t)$ can give the same $H(t)$ (see also the detailed dynamical analysis of Ref.~\cite{Frusciante:2013zop}). In Sec.~\ref{obvia} we outline a strategy to remove  such a degeneracy. The latter can be broken by looking at specific observables in the perturbation sector of the theory which also affects the evolution of the background.

\subsubsection{Perturbation sector}  \label{perturbedd}

The evolution of the  large-scale,  inhomogeneous distribution of matter in the Universe can be computed with a good approximation by using linear cosmological
perturbation theory. The \emph{perturbed sector} of the EFT formalism involves all operators of action~\eqref{example} except $\lambda(t)$.

In the standard gravitational paradigm, and assuming  the quasi-static approximation (see e.g.~\cite{Noller:2013wca} for a throughout discussion)
the evolution of density perturbations $\delta$ is given by 
\begin{equation}\label{linpert}
\ddot \delta + 2 H \dot \delta - 4 \pi G_{N} \rho_m \delta = 0\, .
\end{equation}
This is still true in more general scenarios, at least for those scales in which the quasi-static approximation applies, as long as the Newton gravitational constant $G_N$ is replaced by a more complicated function 
of both time $t$ and comoving Fourier scale $k$ ~\cite{GLPV},
\begin{equation} \label{geff}
G_{\rm eff} \ = \ \frac{1}{8 \pi  \Mb^2(1+\epsilon_4)^2} \ \frac{2 {\cb}  +  \mathring{\mu}_3  - 2 \dot H \epsilon_4 + 2 H \mathring{\epsilon}_ 4 + 2 (\mu + \mathring{\epsilon}_4) ^2 + \ Y_{\rm IR} }{\ 2 {\cb} + \mathring{\mu}_3  - 2 \dot H \epsilon_4 + 2 H \mathring{\epsilon}_ 4 + 2 \dfrac{(\mu + \mathring{\epsilon}_ 4) (\mu - \mu_3)}{1+\epsilon_4} - \dfrac{(\mu - \mu_3)^2}{2 (1+\epsilon_4)^2} +  \ Y_{\rm IR} \ } \ ,
\end{equation}
where we have defined
\begin{align}
\mathring{\mu}_3  \ &\equiv \ \dot \mu_3 + \mu \mu_3 + H \mu_3 ,\\
\mathring{\epsilon}_4  \ &\equiv \ \dot \epsilon_4 + \mu \epsilon_4 + H \epsilon_4\, ,\\
Y_{\rm IR} \ &\equiv \ 3 \left(\frac{a}{k}\right)^2\, \left[2  \dot H \cb - \dot H \mathring{\mu}_3 + \ddot H(\mu - \mu_3) - 2 H \dot H \mu_3 -2 H^2(\mu^2 + \dot \mu)\right]. \label{ir}
\end{align}

An observational probe that is particularly  sensitive  to the specific form of $G_{eff}$ is  the redshifts distortion. Interestingly, 
one could also exploit weak lensing data and constrain an additional observable,  the \emph{gravitational slip parameter} $\gamma_{\rm sl}\equiv \Psi/\Phi$: the ratio of the two gravitational potentials 
defined by the perturbed metric in Newtonian gauge
\begin{equation} \label{newtonian} 
ds^2 = -(1+2\Phi)dt^2 + a^2(t) (1-2 \Psi) \delta_{i j} dx^i dx^j\, .
\end{equation}
Despite the forecasted constraints are less stringent, this observable adds independent lines of evidence and can resolve residual degeneracy in the perturbed sector. 
Again, by specializing the general formulas of Ref.~\cite{GLPV} to action~\eqref{example}, we find
\begin{equation} \label{postn}
1 - \gamma_{\rm sl} \ = \ \frac{(\mu + \mathring{\epsilon}_ 4) (\mu + \mu_3 + 2  \mathring{\epsilon}_ 4) - \epsilon_4 (2 \cb +  \mathring{\mu}_3 - 2 \dot H \epsilon_4 + 2 H \mathring{\epsilon}_ 4)+  \epsilon_4 \cdot Y_{\rm IR}}{ 2 \cb +  \mathring{\mu}_3 - 2 \dot H \epsilon_4 + 2 H \mathring{\epsilon}_ 4 + 2 (\mu + \mathring{\epsilon}_ 4)^2 + Y_{\rm IR}}\, .
\end{equation}

The above defined quantities, $G_{\rm eff}$ and $\gamma_{\rm sl}$, only depend on the three non-minimal coupling functions $\mu$, $\mu_3$ and $\epsilon_4$, which can be taken as appropriate coordinates in the parameter space of modified gravity theories.  The coupling $\mu_2^2$ does not appear in eqs.~\eqref{geff} and~\eqref{postn} but plays a role in the stability and speed of sound of dark energy, to be discussed in Sec.~\ref{stabilityand} below. Moreover, our formalism gives a relatively compact expression for the infra-red, scale dependent term $Y_{\rm IR}$, eq.~\eqref{ir}. Interestingly, such a scale dependence  can be in principle constrained by future data~\cite{Silvestri:2013ne,Amendola:2013qna,Hojjati:2013xqa}. We will further discuss the order of magnitude of the couplings and the consequences on the scale dependence of $G_{\rm eff}$,  in Secs.~\ref{sec:3.3}  and~\ref{sec:5.1} respectively.

\subsection{Theoretical viability and phenomenological constraints} \label{sec:2.2}
We conclude the presentation of the essential features of the EFT formalism by 
discussing  some general conditions that modified gravity models must 
satisfy if they are to be viable.  
With the action written in the standard form~\eqref{example} one can study  the linear dynamics of the propagating scalar degree of freedom once the system has been diagonalized~\cite{GPV,GLPV}. In 
A theory is said to be  sound if such a degree of freedom has neither ghost- nor gradient-instabilities. Besides the stability criteria, in this section we also discuss 
 on the relation between the reduced Planck mass $M_{\rm Pl}=1/\sqrt{8\pi G_N}$ and the present value of the EFT parameter $M^2(t)$, and on the constraints imposed by  nucleosynthesis 
at very early times. 

\subsubsection{Stability and speed of sound} \label{stabilityand}
 
Stability conditions can be analyzed by isolating the scalar propagating degree of freedom contained in the theory and by writing its Lagrangian. 
This is done explicitly in~\cite{GLPV}  by working directly in unitary gauge and using the ADM formalism. 
Equivalently, by a change of coordinates we can make the scalar field's fluctuations reappear explicitly in the theory. By forcing a time diffeomorphism  on the action~\eqref{example}, 
\begin{equation}
t \rightarrow t + \pi(x)
\end{equation}
the spacetime dependent parameter $\pi(x)$ becomes the scalar field fluctuation. The system is then governed by $\pi$ and by the (scalar) metric fluctuations, which can be taken as the gravitational potentials  $\Psi$ and $\Phi$ in the Newtonian gauge~\eqref{newtonian}. The quadratic Lagrangian contains at most one derivative per field. At highest order in derivatives, the presence of the non-minimal coupling functions
$\mu$, $\mu_3$, $\epsilon_4$ produces a mixing between $\pi$ and the gravitational potentials. However, the system can be diagonalized with field redefinitions~\cite{GPV}. Then we are left with the truly propagating degree of freedom $\pi$, decoupled from gravity and governed by the quadratic Lagrangian 
\begin{equation} \label{pai2}
S_\pi = \int \, a^3 \Mb^2 \left[A \left(\mu, \mu_2^2, \mu_3, \epsilon_4\right)\  \dot \pi^2  \ -\  B  \left(\mu, \mu_3, \epsilon_4\right) \frac{(\vec \nabla \pi)^2}{a^2} \right] \,  +\ \dots \, , 
\end{equation}
 where ellipsis stands for terms that are lower in derivatives and the terms $A$ and $B$ read, explicitly, 
\begin{subequations}\label{A-B}
\begin{align} 
A \ &= \ (\cb + 2 \mu_2^2)(1+ \epsilon_4) + \frac34 (\mu-\mu_3)^2 \, , \label{A} \\
B \ &= \ (\cb +  \frac{\mathring{\mu}_3}{2} -  \dot H \epsilon_4 +  H \mathring{\epsilon}_ 4)(1+ \epsilon_4) - (\mu - \mu_3)\left(\frac{\mu - \mu_3}{4(1+ \epsilon_4)} - \mu -  \mathring{\epsilon}_ 4\right)\, . \label{B}
\end{align}
\end{subequations} 
Note that the function $\mu_2$ does not enter $G_{\rm eff}$~\eqref{geff} nor the gravitational slip $\gamma_{\rm sl}$~\eqref{postn}, but does appear in the equation of propagation for $\pi$ and is therefore relevant for the stability of the theory. 

The expressions $A$ and  $B$ must be separately positive. The positivity of $A$  guarantees that there are no ghosts and therefore, ultimately, the soundness of the theory itself (see e.g. the discussion in Ref.~\cite{Cline:2003gs}).
The positivity of term $B$, on the other hand,  enforces the gradient stability condition. This condition prevents 
an instability which is less severe, and it could be relaxed by considering  operators containing higher space derivatives, i.e. terms 
which become important at some high momentum scale $k_{\rm grad}$. 
If such operators appear with the right sign, the exponential growth related to the wrong sign of term $B$ could be limited to the infra-red modes of momenta $k<k_{\rm grad}$.
For a throughout discussion of this issue, and the rather tight observational constraints related to it, we refer the reader to Sec. 2.2 of Ref.~\cite{Creminelli:2008wc}. 
Here, for definiteness, we limit ourselves to the operators displayed in~\eqref{example}, which \emph{do not} contain higher derivatives, 
and thus we will not invoke such a mechanism. In summary, we will simply require both stability conditions 
\begin{subequations} \label{conditions}
\begin{align}\label{noghost}
A \left(\mu, \mu_2^2, \mu_3, \epsilon_4\right) \ &> \ 0 \qquad \qquad \text{no-ghost  condition}\, \\[2mm]
B \left(\mu, \mu_3, \epsilon_4\right) \ &\ge \ 0 \qquad \qquad \text{gradient-stability  condition}\,  \label{gradient}
\end{align}
\end{subequations}
to be independently satisfied. Note also that expression $B$ is proportional to the denominator of $G_{\rm eff}$, eq.~\eqref{geff}. Therefore, requiring that $B>0$ also saves us from a possible pathological behavior of $G_{\rm eff}$.

From~\eqref{pai2}, the propagation speed $c_s$ of dark energy (its ``speed of sound") can be read straightforwardly,
\begin{equation} \label{speed}
c_s^2\ =\ \frac{B}{A}\, .
\end{equation}
Conditions~\eqref{conditions} make $c_s^2$ a positive number as expected, but they are not enough to guarantee that the propagation speed be less than that the speed of light, \emph{i.e.} that $c_s \leq 1$. It has been debated whether or not one should tolerate super-luminal propagation in a low-energy effective theory. Signals traveling faster than light lead to well-known puzzles and paradoxes, such as the presence of boosted reference frames with respect to which such signals arrive before leaving. These macroscopic difficulties might be accompanied with others, more subtle and ``microscopic": 
theories with superluminal propagation have been argued to not admit a consistent ultraviolet completion~\cite{Adams:2006sv}. In our formalism, large values of the structural function $\mu_2^2$ automatically guarantee subluminal propagation, as apparent from eqs.~\eqref{conditions} and~\eqref{speed}. Otherwise, we will generally keep an open-minded attitude on superluminality in the present paper. 
%
%
%
%

\subsubsection{Initial conditions, BBN constraints and screening} \label{icon}
In modified gravity, the attractive interaction  between two gravitating bodies is given by the modified Poisson equation for the gravitational potential $\Phi$
\begin{equation}
-\frac{k^2}{a^2}\Phi=4\pi G_{\rm eff}(t,k)\rho_m \delta_m\, ,
\end{equation}
where $G_{\rm eff}$  can be calculated in the quasi static limit by solving the equations for the metric and for the scalar degree of freedom~\cite{GLPV}, and 
is  explicitly given in eq.~\eqref{geff}. Schematically, it reads
\begin{equation}\label{schematic}
G_{\rm eff}(t)\ = \ \frac{1}{8 \pi \Mb^2(t)} \left [1\, +\, F(\mu,\mu_3,\epsilon_4) \right]\, ,
\end{equation}
where the function $F$ of the couplings and their derivatives vanishes when  $\mu$, $\mu_3$, $\epsilon_4$ go to zero. 
It is important to note that the linear equations of the long-range propagating field contribute the term $F$ inside the square brackets. Indeed, the scalar is not directly coupled to matter in the Jordan frame metric (the one used here), 
but is bound to follow the shape of the Newtonian potential due to its mixing with gravity (see e.g.~\cite{GPV}).  The Einstein frame picture, when available, is even more intuitive, because there the scalar directly couples to matter and mediate a fifth force. 

All such linear effects are constrained in about one part in a thousand by solar system tests. Such a severe bound on the non-minimal couplings $\mu$, $\mu_3$, $\epsilon_4$ would make them completely irrelevant on cosmological scales. Therefore,  we need to assume a  non-linear mechanism of \emph{screening}~\cite{Vainshtein:1972sx,Babichev:2013usa} that suppresses the propagation of the scalar degree of freedom in dense environments or in the vicinity of astrophysical sources.  Here,  and in what follows,  we will be cavalier about this issue and just assume that such a mechanism is at work and it is produced by appropriate higher order (non linear) terms that are not displayed in action~\eqref{example} and/or by the non-linear contributions of the displayed quadratic operators expanded at higher order. This allows us as to contemplate---and constrain with cosmological observations---non-minimal couplings $\mu$, $\mu_3$, $\epsilon_4$  that are \emph{a priori} of order one on cosmic scales (see also discussion at the end of Sec.~\ref{perturbedd}).

The above considerations suggest that the linear effects contained in $F$ in eq.~\eqref{schematic} must be extremely close to zero in the solar system due to screening effects.
But since this is where we measure $G_N$, we conclude that
\begin{equation} \label{refinement}
G_{N}\ \simeq \ \frac{1}{8 \pi  \Mb^2(t_0)} 
\end{equation}
in about one part in a thousand.\footnote{A more precise estimate would be to evaluate $M$ at the ``local" value of the scalar field, say, $\phi_{\rm solar-system}$, which could be different than its cosmological value $\phi_0 \sim t_0$. Such a refinement inevitably involves other details of the theory, such as the precise structure of the operators that are cubic and of higher order in the perturbations and is thus beyond the scope of this paper.} Of course, this is not enough precision for  solar system tests, but it is enough for cosmological observations of large scale structures, and thus for setting our initial conditions at the present time.  
In summary, we assume the following integral relation between $\Mb^2(t)$ and $\mu(t)$,
\begin{equation} \label{integral-mu}
M^2(t)\ =\ M_{Pl}^2 \ e^{\int_{t_0}^{t}\mu(t')dt'}\, .
\end{equation}

Finally, we should mention that there are limits on the possible excursion of the Newton constant from primordial nucleosynthesis to the present time.  Following \emph{e.g.} Ref.~\cite{Rappaport:2007ct}, we will assume that $M^2(t)$ at early times be within a 10\% of its value today, say
\begin{equation} \label{bbncons}
\frac{|M^2(z>10) - M^2_{\rm Pl}|}{M^2_{\rm Pl}} \  \lesssim \ \frac{1}{10}\, .
\end{equation}

\section{Resolving the  background degeneracy}\label{obvia}

A promising starting point to constrain modified gravity models is the neat separation between background and perturbation quantities offered by the EFT formalism. 
However, as we will show below, such a split is not complete yet. Additional analysis is needed if we are  to gain insights from data on viable modified gravity models.

As noted, the expansion history $H(t)$  depends only on the functions $\Mb^2(t)$, $\cb(t)$ and $\Lb(t)$. More precisely, as shown in Sec.~\ref{sec_background},    
these three functions are not independent,  only two of them being sufficient to fully determine $H(t)$. 
However, the converse is not true, in the sense that a given expansion history $H(t)$ does not fully specify $\Mb^2(t)$, $\cb(t)$ and $\Lb(t)$. 
This is related to the fact that a cosmic acceleration mechanism can be provided by either the energy momentum tensor of dark energy,  in virtue of its negative pressure, or by a strong non-minimal coupling to gravity, the 
so called  ``self-acceleration" model. These two limiting cases span a degeneracy in the background sector, that is apparent by looking directly at the modified Friedmann equation
\eqref{frie1}: the dark energy density $\rho_D(t)$ and the function $\Mb^2(t)$ can compensate each other, so that different choices of these functions can produce an identical expansion history $H(t)$. 

Such a degeneracy can be resolved using growth history data, because a time-variation of $\Mb^2(t)$ is always accompanied with a modification of the perturbed sector of the theory. Explicitly, a time-variation of $\Mb^2(t)$  switches on the non-minimal coupling $\mu$ defined in eq.~\eqref{mudef}, which affects the evolution of density inhomogeneities through Eq.~\eqref{geff}.  Note that all the other non-minimal couplings  (such as $\mu_3$ and $\epsilon_4$) are only sensitive to the perturbed sector, and therefore can be fixed \emph{independently} of the information on the cosmic expansion history, 
This is not the case for $\mu$, which cannot be varied arbitrarily without hitting back on the background evolution through~\eqref{mudef}. In this sense, the separation between background and perturbations of the EFT formalism is not complete yet. 

One of the goals of this paper is to show how to efficiently treat the expansion history $H(t)$ and the non-minimal coupling $\mu(t)$ as \emph{independent} quantities and thus finally achieve a complete separation between background and perturbations.  In brief, the strategy is to describe the expansion history with the effective equation of state parameter $\we(t)$ of a \emph{minimally coupled} dark energy model. Then, upon fixing the non-minimal coupling $\mu$ using growth history data, we can back engineer $\Mb^2(t)$ and $\rho_D$ of the full theory.

\subsection{Fiducial background model}

Following standard prescriptions in phenomenological  studies of dark energy, we model the expansion history of the universe by means of an effective equation of state parameter $\we(t)$~\cite{post-fried1,post-fried2}, 
\begin{equation}\label{fiducial}
\bar p_D(t) \ = \ \we(t) \, \bar \rho_D(t) \, .
\end{equation}
In the above, $\we(t)$ is the effective equation of state parameter of a
minimally coupled dark energy model  with pressure and energy density $\bar p_D(t)$ and $\bar \rho_D(t)$ respectively. 
Note that $\we(t)$ is not directly related to the parameters of any fundamental theory; it is just a fitting degree of freedom of the Friedmann model describing the observed  scaling of $\dot{a}/a$.  The resulting best fitting model for the expansion history, hereafter called   {\it fiducial background model},  
 is characterized by the following fraction of non-relativistic matter energy density,
\begin{equation} 
\Oe(y)\ = \ \frac{\bar \om_m^0}{\bar \om_m^0 + (1 - \bar \om_m^0) \, e^{\, -3\!\int_0^y  \, \we(y')  d y' }}\, .
\end{equation}
where $\bar \Omega_m^0$ and $\we(y)$ are the standard Friedmann fitting parameters routinely  constrained via cosmological experiments. 
Note that we have defined 
\begin{equation}
y \ \equiv \ \log\frac{a(t)}{a_0} \ = \ \log\frac{1}{1+z}\, .
\end{equation}
The next step is to find the class of EFT  models that reproduce the same expansion rate $H(t)$.

\subsection{Theories reproducing the same expansion history}
We want to constrain the combined scalings of the coupling $\mu(t)$ and equation of state parameter $w(t)$ (see eq.~\eqref{om0}) which reproduce the fiducial model~\eqref{fiducial}.
To this pourpose, let us start by defining the matter density parameter $\Omega_m$ as   
\begin{equation}\label{om}
\om_m(y)\, = \, \frac{\rho_m(y)}{\rho_m(y) + \rho_D(y)}\, .
\end{equation}
Since the fiducial background model is minimally coupled, we require that the expansion rate $H$ of each EFT representative 
be identical to 
\begin{equation} \label{h-fiducial}
 H^2  \  = \  \frac1{3 M_{\rm Pl}^2} (\bar \rho_m + \bar \rho_{D})\, 
 \end{equation}
that is to the expansion rate of  a Friedmann model augmented by a minimally coupled dark energy component. (We are here  assuming  that 
observations are well described in terms of this effective model.)
Note, also, that one can allow the fiducial background to have an energy density of matter $\bar \rho_m$ different than the physical one, $\rho_m$. However, since they both scale as $a^{-3}$, they can differ 
at most by a constant  factor $\rho_m = \kappa \bar \rho_m$. 

By imposing the equality between the right hand sides of~\eqref{frie1} and~\eqref{h-fiducial} we obtain
\begin{equation}\label{itself}
 \kappa \, \Oe (y) \, M_{\rm Pl}^2 \ =\  \Oee(y)\, \Mb^2(y)\, .
\end{equation} 
This identity links the functional space of  theories, described via the functions $w(t)$ and $\Mb^2(t)$, to a given expansion history, whose information is 
 condensed in the discrete parameter $\bar \Omega_m^0$ and the function $\we(t)$. 
By deriving the above and after some straightforward algebra, we obtain the relation
\begin{equation}\label{eq:om_om_bar}
\we(1 - \Oe) = w(1 - \om_m),
\end{equation}
that can be used to express $\om_m$ in terms of $w$. 

By~\eqref{itself} and~\eqref{eq:om_om_bar} we get 
\begin{equation} \label{Msquared}
M^2(y) \ = \ \kappa \, M^2_{\rm Pl} \,  \frac{\Oe w }{w - \we (1- \Oe )}\, .
\end{equation}
and, by exploiting the initial  condition  $M(t_0) = M_{\rm Pl}$  (cfr. eq.~\eqref{refinement}), 
we deduce that the parameter $\kappa$ regulates, at the same time, the ratio between $\Omega_m^0$ and $\bar \Omega_m^0$ and that between $M^2$ and $M_{\rm Pl}^2$ during matter domination:
\begin{equation}
\kappa \ = \ \frac{\Omega_m^0}{\bar \Omega_m^0}\, , \qquad \qquad M^2(t \rightarrow 0)\ = \ \kappa \, M^2_{\rm Pl}\, .
\end{equation}
Because the limits from nucleosynthesis constrain the total displacement of $M^2$ from the Planck value to  about 10\%~\eqref{bbncons}, we simply  set $\kappa = 1$ (and thus $\Omega_m^0 = \bar \Omega_m^0$) from now on. 

By using~\eqref{eq:om_om_bar} we can now directly calculate the derivative of $\Mb^2(t)$, and thus the coupling $\mu$. We obtain
\begin{equation} \label{represennn}
\mu \ = \ \frac{H (1 - \Oe)}{w - \we + \we \, \Oe}  \left[3\we \left( w -\we \right)  +  \frac{d \we}{d y} - \frac{\we}{w} \, \frac{d w}{d y} \right]\, .
\end{equation}
with initial condition $w(0) = \we(0)$. 
Note that if  $\mu = 0$  one recovers the fiducial background model, $w = \we$, i.e. dark energy is correctly modeled in terms of a minimally 
coupled (``quintessence"-like) scalar degree of freedom.  

To summarize, if an EFT  theory reproduces the expansion history given in terms of the phenomenological Friedmann parameters $\Oe(t)$ and $\we(t)$,  then the background functions  
 $w(y)$ and $\mu(y)$ characterizing that theory need to satisfy the constraint \eqref{represennn}. Note that the background sector of EFT models is now completely specified since 
 the additional coefficients $\cb$ and $\lambda$ are automatically computed once $w$ and $\mu$ are known by means of eqs.~\eqref{c2} and~\eqref{L2}.

\subsection{Dimensionless EFT couplings} \label{sec:3.3}

At this stage it is worth making a few general considerations on the size of the 
non-minimal couplings $\mu$, $\mu_2^2$, $\mu_3$, $\epsilon_4$. Their dimensions  are naturally set by the appropriate powers of the Hubble parameter $H$, as suggested by inspecting  the  action~\eqref{example}. 
Clearly, dimensional analysis alone cannot set the amplitude of the couplings. Consider, for example, $\mu$ defined in eq.~\eqref{mudef}. If $\mu \ll H$, the value of $M^2(t)$ barely changes on cosmic time scales. As a consequence (see Eq.~\eqref{frie1})  cosmic acceleration is generated  by a negative pressure component, and not by a genuine self-acceleration/modified gravity effect. 
Incidentally, this situation has a most notable example in $f(R)$ gravity~\cite{DeFelice:2010aj}. It is now understood~\cite{Wang:2012kj} that the observational limits on $\mu$ for $f(R)$ theories are rather strict---of order $\mu \lesssim 10^{-3} H$. This is because the chameleon mechanism~\cite{Khoury:2003aq}, on which $f(R)$ theories rely, is scarcely efficient at screening the unobserved effects of modified gravity from the solar system environment. 
Such a small value of $\mu$, although perfectly legitimate, relegates the scalar field to the role of a spectator on the largest cosmological scales,  and a ``standard" cosmological constant term is still required to drive cosmic acceleration. 

In this paper, on the contrary, we focus on those  scenarios in which the scalar field and the modified gravity mechanism do play the main role in the present cosmic acceleration. We will thus consider, \emph{a priori},  non-minimal couplings of order one in Hubble units\footnote{This assumption, in general, has consequences also on the scale dependence of $G_{\rm eff}$ and $\gamma_{\rm sl}$. We comment on this later on in Sec.~\ref{sec:5.1}}, with no particular hierarchy among them. This also means that, as already noted in Sec.~\ref{icon}, we need to assume a mechanism of screening other than---and more powerful of---the chameleon, in order to make such large couplings compatible with the physics of the solar system. 

Finally, we comment on the time-dependence of the couplings. Despite the formalism is general enough to allow for any time dependence, 
 it is natural to assume that  modified gravity effects become important only at late epochs, when dark energy dominates. This is, in particular,  a characteristic  feature of 
 explicit scalar tensor models of gravity. Indeed, when remapping the EFT lagrangian back into the usual  covariant form, the EFT couplings  typically contain one or more time derivatives of a suitably normalized scalar field $\phi$ (see \emph{e.g.} App. C of Ref.~\cite{GLPV}). As expected, these functions  become subdominant at early time,  during matter domination, when the energy density of $\phi$ is negligible. In doing so,  they do not modify the successful predictions of the standard model of cosmology at early epochs.

The above assumptions translate into the following definitions of the dimensionless coupling functions $\eta$-$\eta_i$ operating in the perturbation sector,
\begin{subequations}\label{repback}
\begin{align} 
\mu \ &= \ \eta\, H \, (1-\Oe) \, , \label{mupar}\\
\mu_2^2 \ &= \ \eta_2\, H^2 \, (1-\Oe) \, \label{eta2},\\
\mu_3 \ &= \ \eta_3\, H \, (1-\Oe) \, \label{eta3},\\
\epsilon_4 \ &= \ \eta_4 \, (1-\Oe) \, . \label{eta4}
\end{align}
\end{subequations}

Note that all relevant background quantities can be calculated once $\we$ and $\eta$ are assigned. For instance, let us consider  $\cb$, the crucial term for assessing  the stability of the modified gravity theories (see  next section). From  $\we$ and $\eta$ we can calculate $w$ by solving the following equation [that descends straightforwardly from~\eqref{represennn}]
\begin{equation} \label{represen2}
\eta \ = \ \frac{1}{w - \we + \we \, \Oe}  \left[3\we \left( w -\we \right)  +  \frac{d \we}{d y} - \frac{\we}{w} \, \frac{d w}{d y} \right]\, ,
\end{equation}
together with the initial condition $w(0) = \we(0)$. Then the amplitude of $\cb$ follows from equation~\eqref{c2} [see also eq.~\eqref{ccc} below].

 In Table~\ref{symbols} we summarize the independent dimensionless functions of our formalism (central column) and relate them to the original coefficients in action~\eqref{example} (left column) and to their observational effects (right column).

 \renewcommand{\arraystretch}{1.6}

\begin{savenotes}
\begin{table}
\centering
    \begin{tabular}{|l||c|l|}
 \hline
Time-dependent \ \ & Dimensionless & \\[-3mm]
 \text{couplings in action~\eqref{example}} & 
  \text{free functions} & Estimators
    \\ \hline
       \hline
         \multirow{2}{*}{ $\Lb(t)$, \ $\cb(t)$, \ $\mu(t) \equiv \frac{dM^2(t)}{dt}$\ \  }  & $\we(y)$  &  Expansion history  \\ \cline{2-3} 
           & $\eta(y)$     &  \multirow{3}{*}{Growth rate} \\ \cline{1-2} \cline{1-1}
    $\mu_3(t)$  & $\eta_3(y)$  &\\ \cline{1-2} 
       $\epsilon_4(t)$ & $\eta_4(y)$   & \\ \hline
            $\mu_2^2(t)$  & $\eta_2(y)$   & Unconstrained\footnote{The function $\eta_2$ enters the sound speed of dark energy~\eqref{speed}. Therefore, a relevant exception to this statement is when $\mu_2$ is \emph{much larger} than all other couplings. In this case, the sound speed of dark energy is practically zero and we are in the presence of a ``clustering quintessence" type behavior (see e.g.~\cite{Sefusatti:2011cm}). }  (but relevant for stability) \\ \hline
  \end{tabular}
  \normalsize
  \caption{In the central column we display the free independent functions of our formalism. A complete separation between background expansion history (the function $\we$) and perturbations (the functions $\eta$-$\eta_i$) is now achieved. Only $\eta$, $\eta_3$ and $\eta_4$ are true non-minimal couplings entering $G_{\rm eff}$ [eq.~\eqref{geff}] and the gravitational slip $\gamma_{\rm sl}$ [eq.~\eqref{postn}]. The left column displays the associated coefficients of action~\eqref{example}.}
\label{symbols}
\end{table}
\end{savenotes}

\section{Exploring the space of theories} \label{sec:4}

In order to  explore the space of  theories and confront  predictions with observations, we need to chose a particular  form for the free functions $\we(y)$ and $\eta(y)$-$\eta_i(y)$. We will start discussing the former, that we set to a constant from now on,  although different choices are possible given the absolute generality of the  formalism presented in Sec.~\eqref{obvia}. We then discuss a convenient ansatz for the function $\eta$, that will produce closed exact expressions for all the relevant quantities of the formalism. As for the remaining couplings $\eta_i$,  we explore the easiest ansatz: the constant one. The main goal  is to determine the region of stability of
modified gravity  theories in the $\eta_i$ space for different values of $\we$. 

\subsection{The case for a constant $\we$}

A large body of observations  currently suggest that a single fitting  degree of freedom, i.e. a  constant $\bar w$, 
is more than enough for describing the cosmic  expansion rate  $H$ with high precision. 
More elaborated extensions of the Friedman paradigm, as those obtained by allowing an explicit time dependence of   the  effective equation of state $\bar w(t)$ 
do not provide higher resolution insights into the expansion history of the universe. As a matter of fact,  we have verified that 
the minimum $\chi^2_{\nu}$ value obtained from the analysis of the Hubble diagram of Supernovae Ia collected in the  Union 2 data set,  a state of the art
compilation of data for  SNIa in the redshift range $0 < z < 1.4$ \cite{amanullah}, is nearly similar $(\chi^2_{\nu, min} \sim 562$, for $\nu \sim  580$ degrees of freedom)  irrespective of 
whether the data are fitted assuming a constant model for $w$ or a time evolving model of the type $w_0+w_a (1-a)$. Therefore, from now on,  we will set 
\begin{equation} \label{choicew}
\we(t) \ = \ \we\, .
\end{equation}
Clearly, new and more precise observations will likely impose the adoption of a more refined  fitting scheme. Notwithstanding, all our arguments 
can be generalized in a straightforward way if  bayesian evidence should point toward the necessity of parameterizing 
$\bar w$ with more than 1 parameter. The choice~\eqref{choicew} obviously contains the $\Lambda$CDM background behavior as a limit, and easily allows to consider small deviations from it, which is one of our main targets.

The fractional energy density of non-relativistic matter of the fiducial background model, $\Oe$, proves a useful cosmological ``clock" for late time cosmology because it naturally triggers dark energy related effects. 
In fact,  $\Oe$ interpolates smoothly between the matter domination epoch, when it evaluates $1$, and the  dark energy domination epoch, when it evaluates $\sim 1/3$. Since we will often use this variable in the following, 
we simply label it as $x$:  
\begin{equation} \label{oe2}
x \ \equiv \ \Oe(\we= {\rm const.})\ = \ \frac{\om_m^0}{\om_m^0 + (1 - \om_m^0) \, e^{-3  \we  y }}\, .
\end{equation}
Derivatives with respect to $y$ and with respect to $x$ are related by
\begin{equation} \label{derivatives}
\frac{d}{dy} \ = \ 3 \we \x (1-\x) \, \frac{d \, \, }{d \x}\, .
\end{equation}
Also, from now on, derivatives with respect to $\x$ will be indicated with a prime symbol $'$.  Note that, 
as a consequence of~\eqref{choicew}, and by using $x$ as a ``time" variable, eq.~\eqref{represen2} reduces to
\begin{equation} \label{etaaa}
\eta(x)\ =\ \frac{3 \we}{w(x)-\we + \we \x} \left[w(x) - \we - \we \x \, (1 - \x)\frac{w'(x)}{w(x)} \right]\, .
\end{equation}

\subsection{The parameterization of the couplings}

Here we discuss how we concretely  parameterize the---so far, completely general---time dependent couplings $\eta, \eta_2, \eta_3$ and $\eta_4$ in terms of a suitable number of real coefficients.
Most of this Section will be devoted to the ``Brans-Dicke" or ``background" sector of the theory, the one obtained by setting the coefficients of all quadratic operators, $\eta_2$, $\eta_3$ and $\eta_4$, to zero. In this limited sector, the issue of stability is easily addressed.  If we set to zero  the high oder couplings in eq.~\eqref{A-B},  the terms $A$ and $B$ become identical,   and the stability conditions Eqs.~\eqref{noghost} and~\eqref{gradient} coincide. The stability criterion  boil down to r
\begin{equation} \label{stabbrans}
\cb + \frac{3 \mu^2}{4} \ > \ 0\, .
\end{equation}
By using eqs.~\eqref{c2},~\eqref{om} and \eqref{mupar} we obtain 
\begin{equation} \label{ccc}
\cb = \frac{H^2 (1- \x)}{2} \left[ 3 \we \, \frac{1 +w}{w} + (5+ 3\we + 3\we \x) \frac{\eta}{2}  - (1-\x)\eta^2 - 3 \we \x (1- \x)\eta' \right]\, ,
\end{equation}
where $w$ is a solution of~\eqref{etaaa} with the initial condition $w(\x = \Omega_m^0) = \we$. Note that,  in the minimally coupled case ($\eta = 0$, $w = \we$),  the only non-vanishing term inside the square brackets is the first one, and $\cb$ reduces to $(\rho_D + p_D)/( 2 M_{\rm Pl}^2)$~\cite{Creminelli:2008wc}.
Analogously, we can calculate the $\lambda$ background function,
\begin{equation} \label{lambda}
\lambda = \frac{H^2 (1- \x)}{2} \left[ 3 \we \, \frac{1 - w}{w} + (7- 3\we - 3\we \x) \frac{\eta}{2}  + (1-\x)\eta^2 + 3 \we \x (1- \x)\eta' \right]\, .
\end{equation}
Thanks to the specific recasting  of the non-minimal coupling $\mu$ [cfr. eq.~\eqref{mupar}] an overall factor of $H^2 (1-x)$ collects from the stability  condition~\eqref{stabbrans}, which then becomes\footnote{Strictly speaking, in the absence of \emph{any} higher derivative operator, fifth force experiments would already confine 
the size of the remaining non minimal coupling $\eta$ to completely irrelevant and uninteresting values. This is because, if no higher derivative operators are present, we can only confide in the chameleon mechanism~\cite{Khoury:2003aq} to screen the unobserved modified gravity effects. Such a mechanism is known (see~\cite{Wang:2012kj}) to be not enough powerful to produce an appreciable amount of self-acceleration and, at the same time, be compatible with solar system tests. Nonetheless, here and in the following  we contemplate order one values of the $\eta$ parameter even just in the background sector, which constitutes the skeleton of our formalism. The tacit assumption being that, if the remaining $\eta_i$ are absent, higher derivatives \emph{cubic} operators beyond those displayed in action~\eqref{example} will produce sufficient ``Vainshtein"~\cite{Vainshtein:1972sx,Babichev:2013usa} screening.} 
\begin{equation} \label{conditionnn}
3 \we \, \frac{1 +w}{w} + (5+ 3\we + 3\we \x) \frac{\eta}{2}  +\frac12 (1-\x)\eta^2 - 3 \we \x (1- \x)\eta' \ >\ 0 \, .
\end{equation}

To proceed further and extract information from growth history data, we need to supply a specific parametric form for the non-minimal coupling parameter $\eta(x)$.  
This is a critical step, since it involves the incorporation of model dependent assumptions into the EFT formalism.
Ideally, the chosen model should have a minimal impact on our exploration strategy, that is, it should not severely restrict 
the general space of theories  to which the EFT formalism give access. In this sense, a  most generic and uninformative ansatz  is of the type 
\begin{equation} \label{verygeneral}
\eta = A \x^{-1} + B + C \x + D \x^2\, .
\end{equation}
In the absence of any theoretical prior,  there are only a couple of practical \emph{desiderata}  that  help in  tuning the coefficients of~\eqref{verygeneral}. It would be useful
\begin{enumerate}
\item to have an exact solution of~\eqref{etaaa} for $w$ that can be written in closed form.
\item to have condition~\eqref{conditionnn} satisfied for some values of our parameters even when $\we$ drops below $-1$. i.e. to maximize the region of the parameter space which represents  
a well-behaved theory.
\end{enumerate}

We find that the above requirements are met by specializing~\eqref{verygeneral} to the two-parameter expression
\begin{equation}\label{ansatz}
\eta (\x) \ = \, (\bb-\aaa) \frac{x_0}{\x} \, +\,  [\aaa - \bb (2 + x_0)] \, \x \, +\, 2 \bb\, \x^2\, ,
\end{equation}
where we have defined $x_0 \equiv \Oee^0$.
Such an ansatz allows a rather simple closed expression for the functions  $w$,
\begin{equation} \label{w}
w \ = \ \we\ \frac{1- \x }{1- \x\, \exp\left[\frac{(\aaa -\bb + \bb x) (\x - x_0)(1 - \x)}{3 \we \x}\right]}\, ,
\end{equation}
and $\Oee$
\begin{equation}\label{omeg}
\Oee=\x \exp\left[ \frac{\left( \aaa-\bb+\bb x \right ) \left (1-\x \right )\left (\x-x_0\right )}{3\we x} \right ]\, ,
\end{equation}
as required by point 1 above. 
The issue of stability for the background sector (point 2) is discussed throughout in the next subsection.

\begin{table}
\centering
    \begin{tabular}{|l||c|l|}
 \hline
Time-dependent \ \ & Dimensionless & \\[-3mm]
 \text{couplings in action~\eqref{example}} & 
  \text{free functions} & Chosen parameterization
    \\ \hline
       \hline
         \multirow{2}{*}{ $\Lb(t)$, \ $\cb(t)$, \ $\mu(t) \equiv \frac{dM^2(t)}{dt}$\ \  }  & $\we(x)$  &  $= \, \we \qquad \qquad (const.)$  \\ \cline{2-3}
                    & $\eta(x)$     & $= \, (\bb-\aaa) \frac{x_0}{\x} \, +\,  [\aaa - \bb (2 + x_0)] \, \x \, +\, 2 \bb\, \x^2$  \\ \hline
    $\mu_3(t)$  & $\eta_3(x)$  & $ = \, \eta_3 \qquad \qquad (const.)$ \\ \hline
       $\epsilon_4(t)$ & $\eta_4(x)$   & $ = \, \eta_4 \qquad \qquad (const.)$ \\ \hline
            $\mu_2^2(t)$  & $\eta_2(x)$   & $ = \, \eta_ 2 \qquad \qquad (const.)$ \\ \hline
  \end{tabular}
  \normalsize
  \caption{Attached to the same setup of Table~\ref{symbols} we display in the right column the explicit parameterization with which we explore the space of theories}
\label{symbols2}
\end{table}

Finally, we can now complete the parameterization scheme by exploiting the advantages provided by the EFT formalism in the perturbation sector. Since the functions $\eta_2(x)$, $\eta_3(x)$ and $\eta_4(x)$ do not affect the dynamics of the background, they are essentially unconstrained and they can be chosen arbitrarily. We will thus make the simplest ansatz,  the constant one:  $\eta_2(x) = \eta_2$, $\eta_3(x)=\eta_3$ and $\eta_4(x)=\eta_4$.

\subsection{Stability in theory space} \label{sec:4.3}
When dealing directly with a covariant Lagrangian, one has to make sure that the background solution of interest is stable under small fluctuations. This analysis cannot be performed without explicitly calculating the 
evolution equations for the background.  On the opposite, the EFT formalism allows to by-pass this lengthy calculation by reducing the issue to solving the algebraic inequalities~\eqref{conditions} containing  the non-minimal couplings $\eta$-$\eta_i$, their time derivatives and the effective equation of state $\we$. 
The quantities $A$ and $B$ defined in~\eqref{A-B}, which need to be separately positive, can be straightforwardly calculated in terms of our set of parameters $\we, \alpha, \beta, \eta_2, \eta_3, \eta_4$ using eqs.~\eqref{repback}, \eqref{ccc}, \eqref{ansatz}, and~\eqref{w}. We do not quote them explicitly because their  expressions are rather involved. 

\begin{figure}[t] \vspace{-1cm}
\centering
\includegraphics[trim = 2.75cm 6.5cm 2.75cm 6.5cm, scale=0.40]{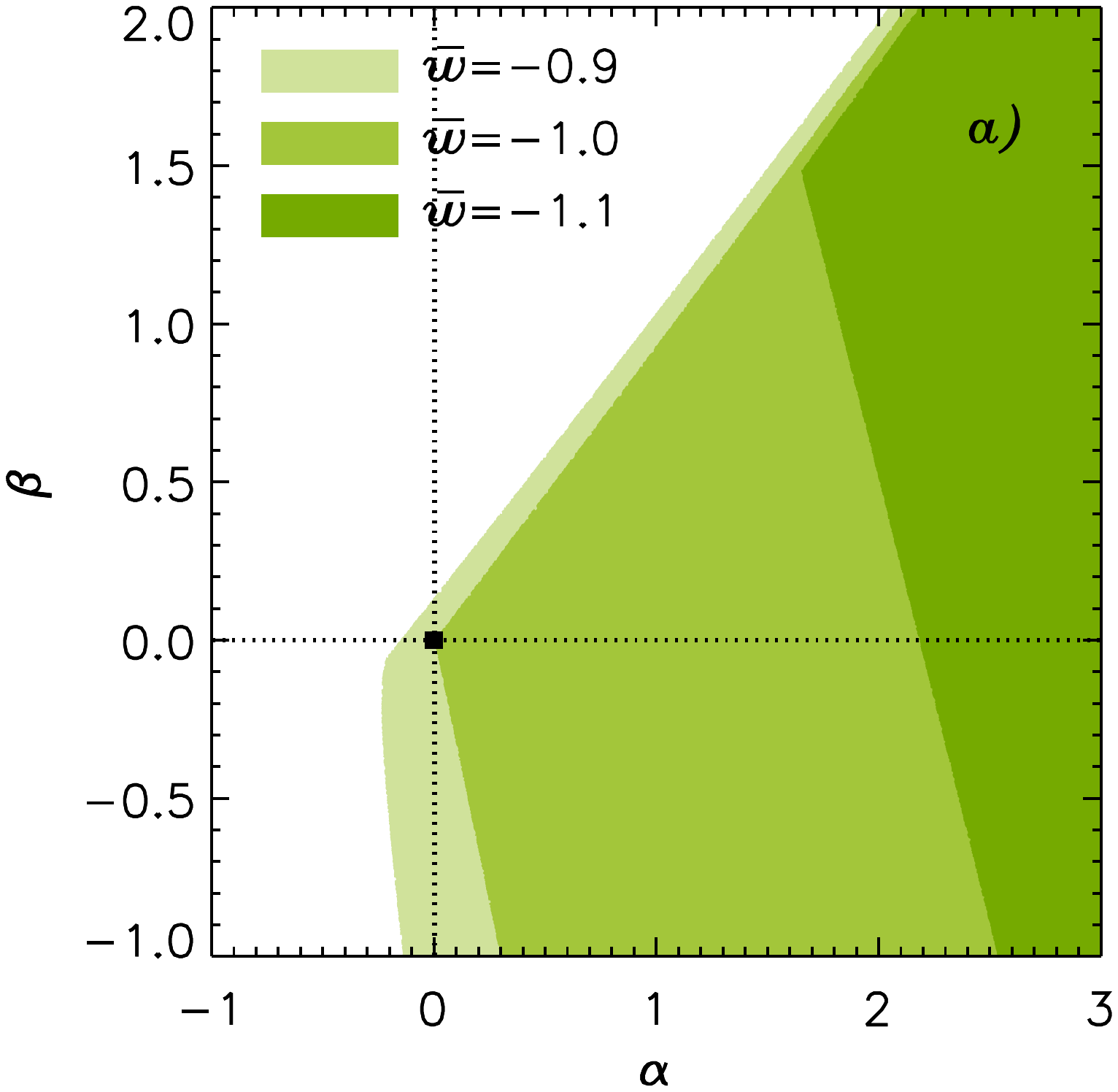}
\includegraphics[trim = 2.75cm 6.5cm 2.75cm 6.5cm, scale=0.40]{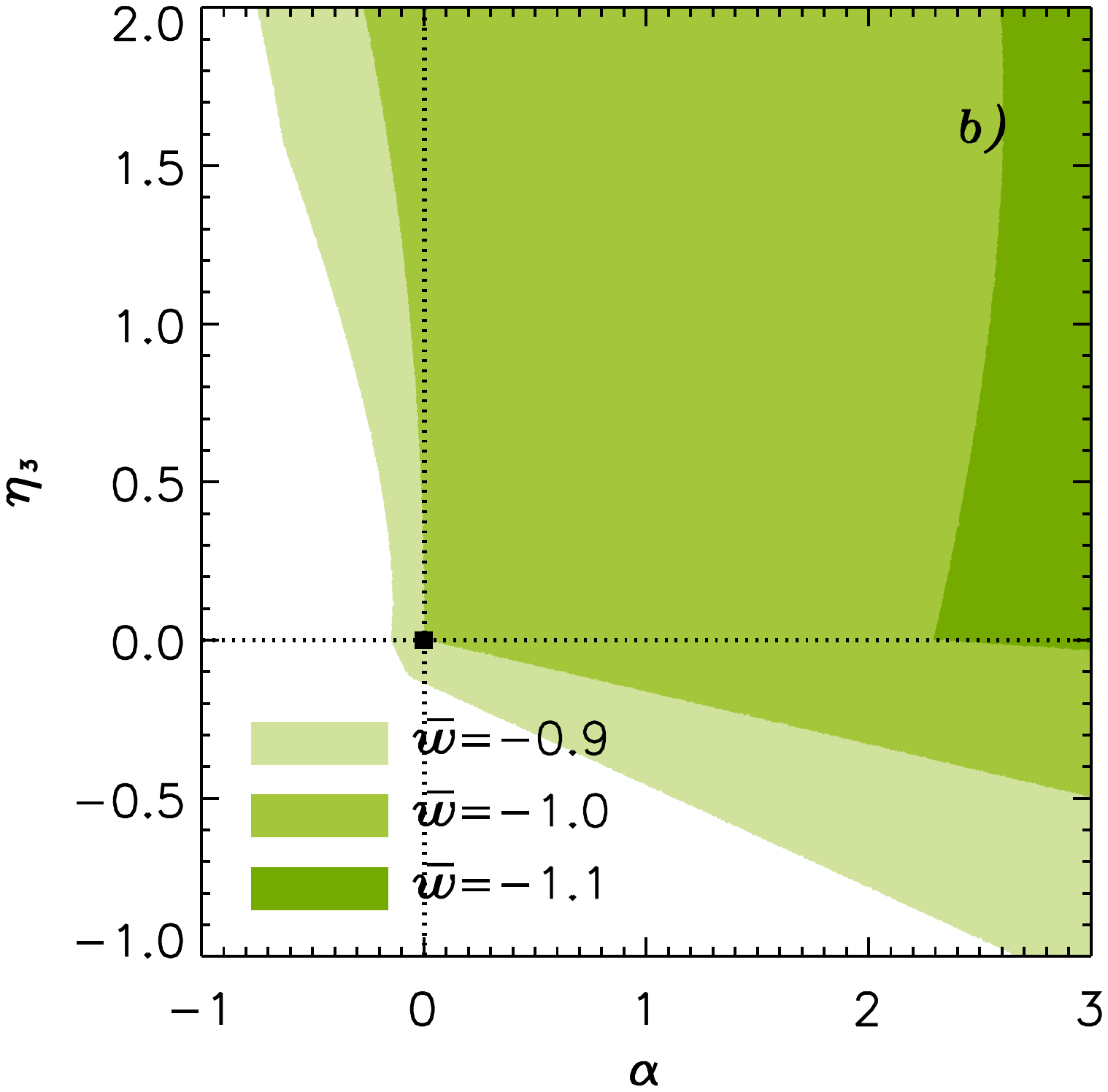} 
\includegraphics[trim = 2.75cm 6.5cm 2.75cm 6.5cm, scale=0.40]{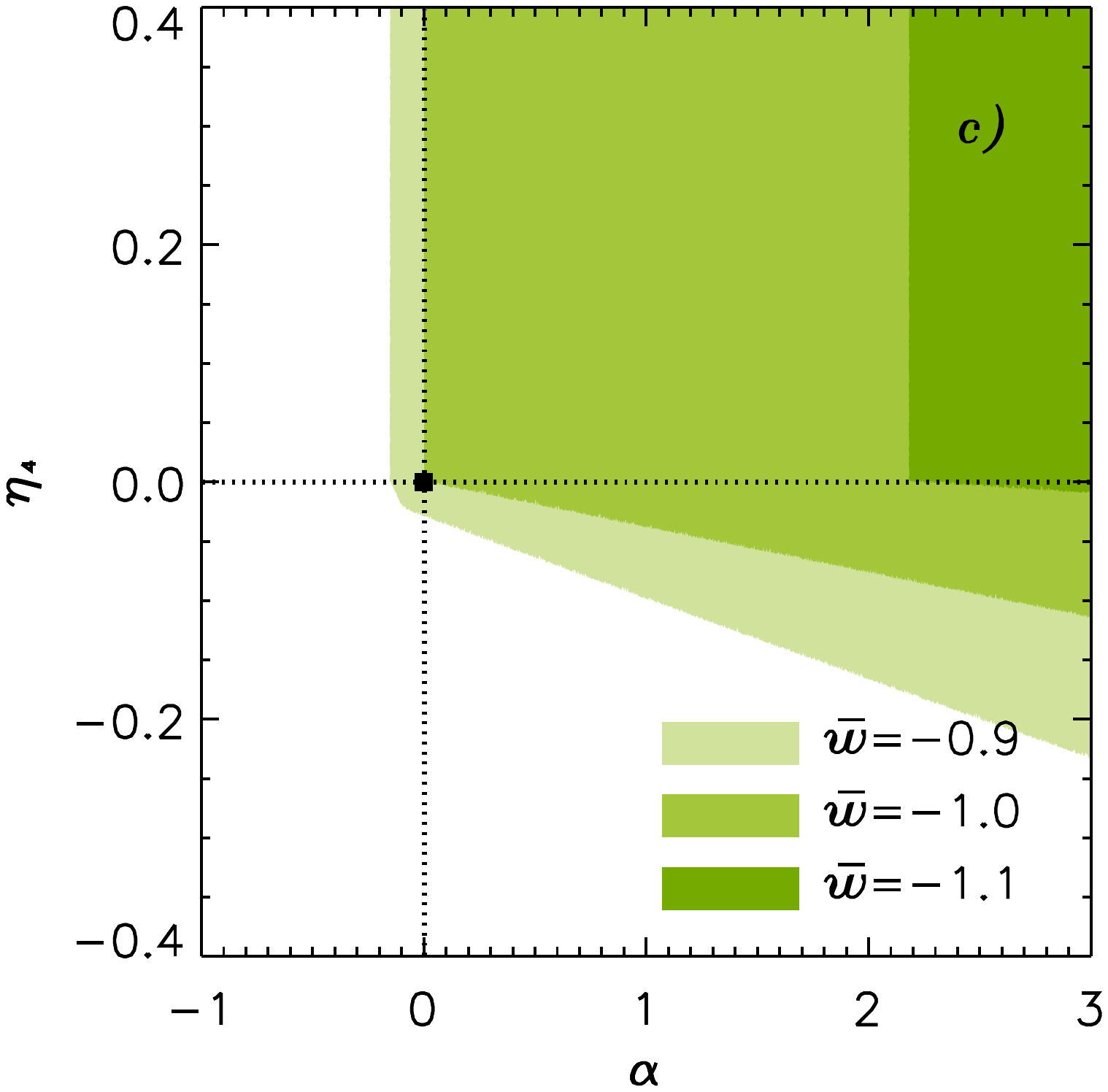}
\includegraphics[trim = 2.75cm 6.5cm 2.75cm 6.5cm, scale=0.40]{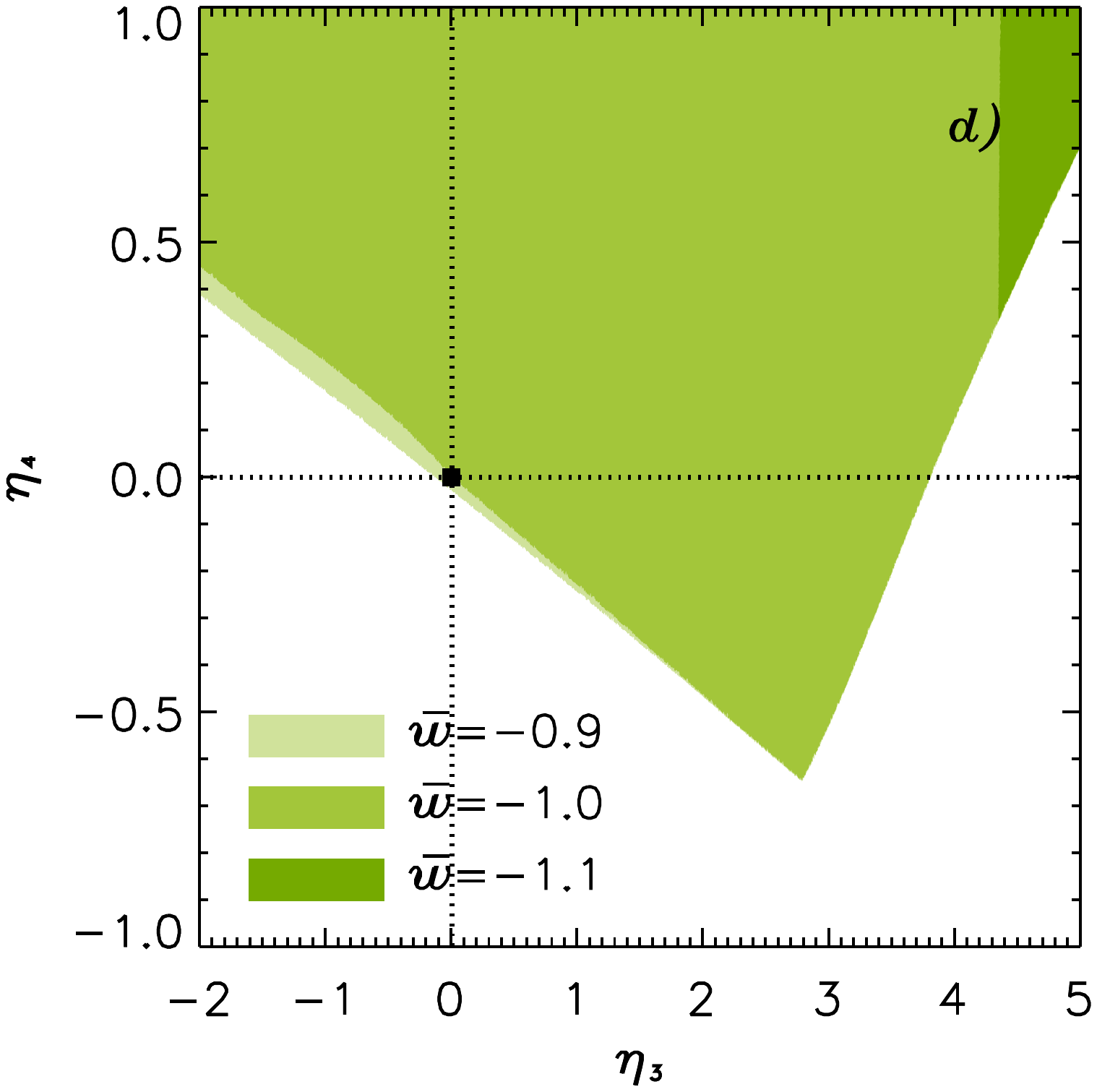} \vspace{0.5cm}
\caption{\label{fig:stabeta20}  
The region of stability in the planes $\alpha$-$\beta$, $\alpha$-$\eta_3$, $\alpha$-$\eta_4$ and $\eta_3$-$\eta_4$ planes. Different shaded area corresponds to different choices of the effective equation of state parameter $\we$, that is  $-1.1, -1$, and $-0.9$. In all the panels we have set $\eta_2$=0.  The parameters $\alpha$, $\beta$, $\eta_3$ and $\eta_4$ not directly shown in the panels are set to zero. General Relativity is the full black square. }  
\end{figure}

\begin{figure}[t] \vspace{-1cm}
\centering
\includegraphics[trim = 2.75cm 6.5cm 2.75cm 6.5cm, scale=0.40]{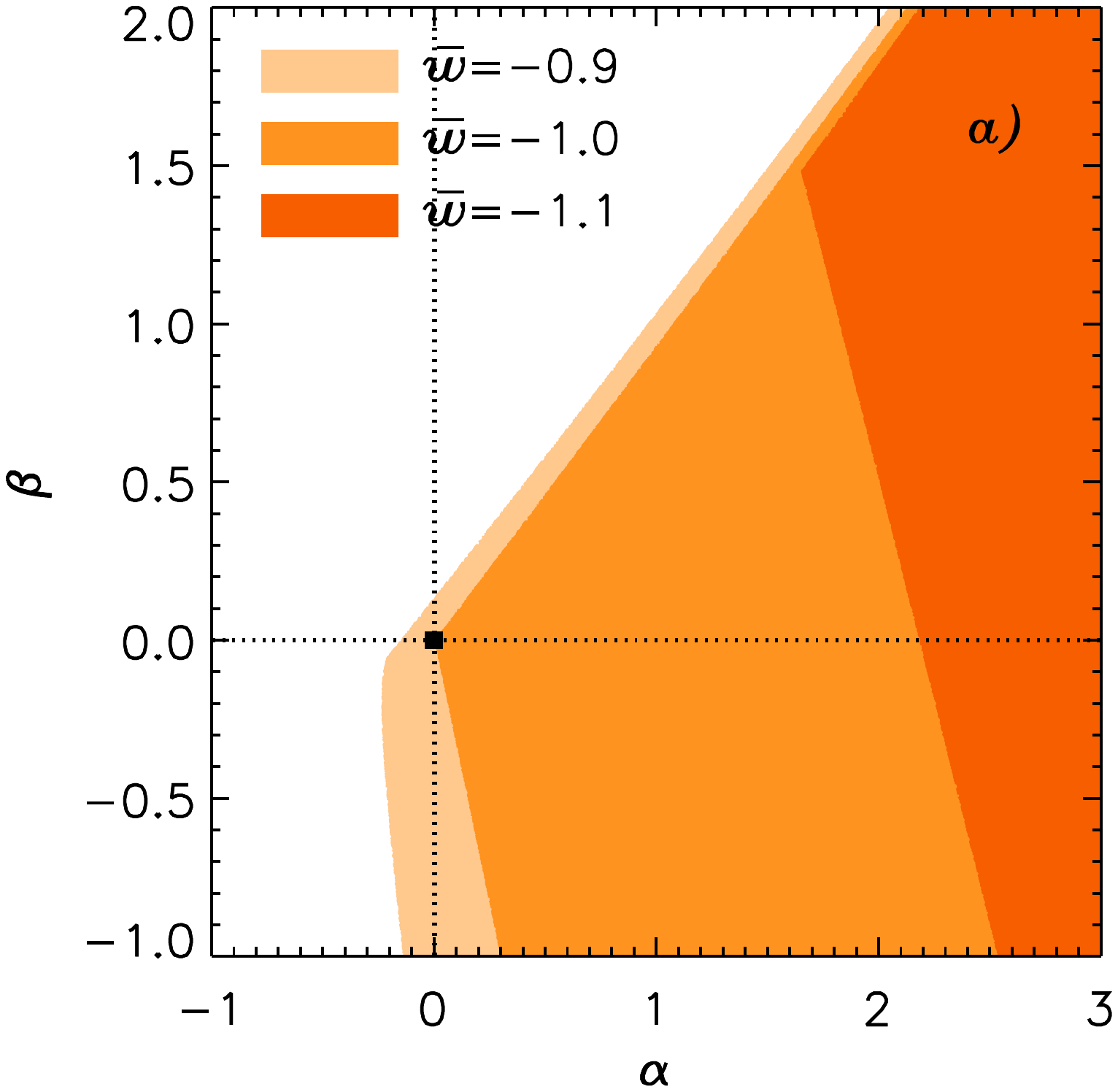}
\includegraphics[trim = 2.75cm 6.5cm 2.75cm 6.5cm, scale=0.40]{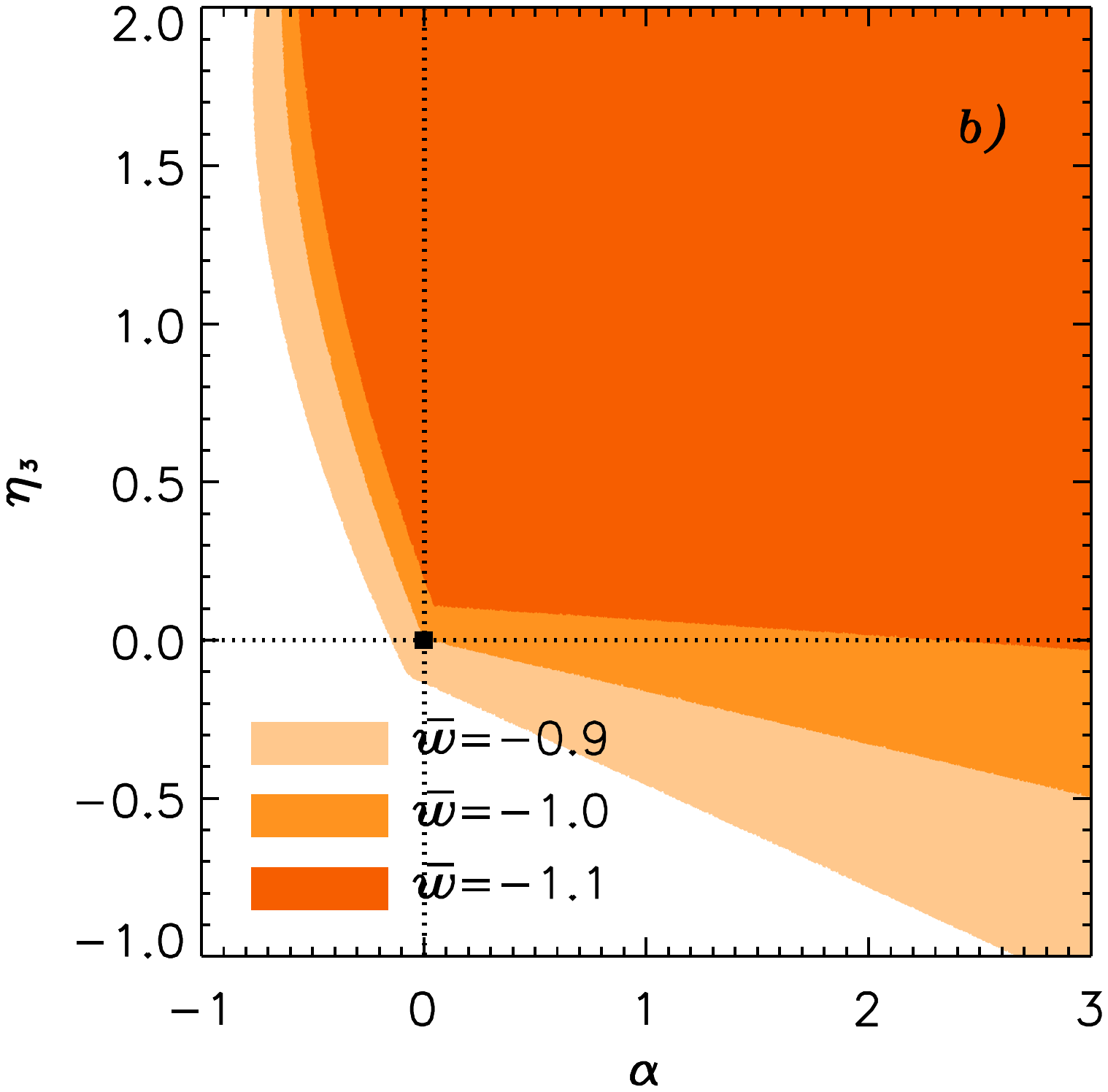} 
\includegraphics[trim = 2.75cm 6.5cm 2.75cm 6.5cm, scale=0.40]{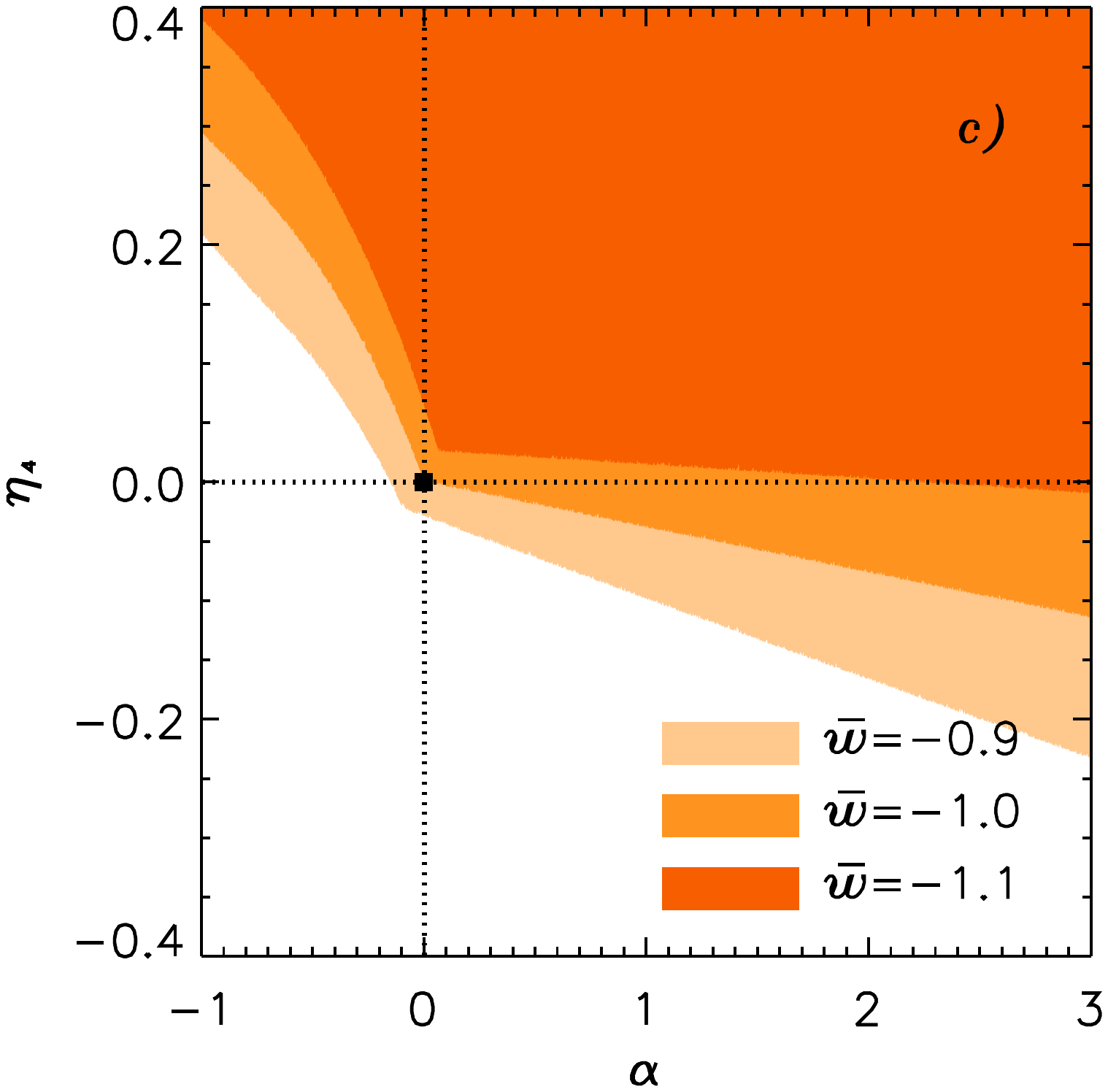}
\includegraphics[trim = 2.75cm 6.5cm 2.75cm 6.5cm, scale=0.40]{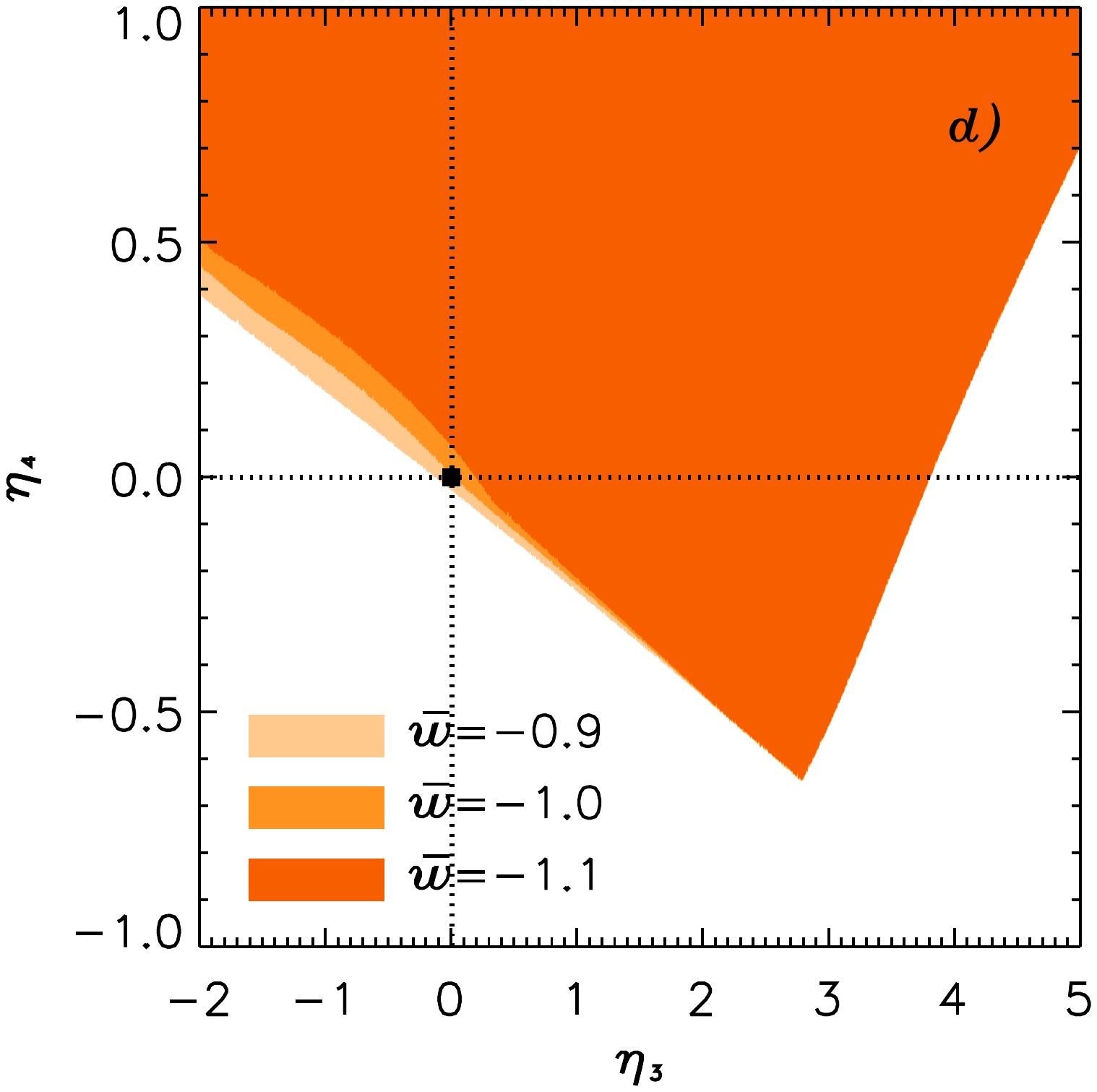} \vspace{0.5cm}
\caption{\label{fig:stabeta2}  
The region of stability in the planes $\alpha$-$\beta$, $\alpha$-$\eta_3$, $\alpha$-$\eta_4$ and $\eta_3$-$\eta_4$ planes. Different shaded area corresponds to different choices of the   
effective equation of state parameter $\we$, that is  $-1.1, -1$, and $-0.9$. In all the panels we have set $\eta_2=10^{6}$.  The parameters $\alpha$, $\beta$, $\eta_3$ and $\eta_4$ not directly shown in the panels are set to zero. General Relativity is the full black square. }  
\end{figure}

The Brans-Dicke sector of the theory ($\eta_2=\eta_3=\eta_4=0$) is spanned in our approach by the two parameters $\alpha$ and $\beta$. However, when we turn to other directions and switch on also $\eta_3$ and/or $\eta_4$, it looks more economical to leave $\alpha$ alone as a measure of the Brans-Dicke coupling, effectively setting $\beta = 0$. The results in terms of the stability of the theories are illustrated in Figures~\ref{fig:stabeta20} and~\ref{fig:stabeta2}. The stability regions are derived by imposing conditions~\eqref{conditions} at all cosmic epochs. The difference between the two Figures is the assumed value of $\eta_2$. Such a parameter can play a relevant role in the no-ghost condition~\eqref{noghost}. We thus consider two limiting cases. In Figure~\ref{fig:stabeta20} we assume a negligible value for $\eta_2$, whereas in Figure~\ref{fig:stabeta2} we consider the opposite limit, and help the stability by turning on a large value of $\eta_2$. Effectively, Figure~\ref{fig:stabeta2} shows the gradient stability regions only.

A universal---and rather expected---feature emerging from our plots is that the region of stability shrinks when $\we$ decreases. 
As a by-product, we can address in full generality the following theoretical problem: under which conditions is it possible to have \emph{stable violations of the null energy condition (NEC)}, and thus an effective equation of state $\we < -1$? Naively, one might think to achieve this with a minimally coupled scalar field, just by flipping the sign of its kinetic term~\cite{Caldwell:1999ew}. 
However,  this turns out to be catastrophic since the considered  theory  would inevitably develop ghost instabilities~\cite{Cline:2003gs}. The fact that a minimally coupled scalar \emph{cannot} produce a super-accelerating equation of state is clearly visible in Figures \ref{fig:stabeta20}   and \ref{fig:stabeta2}. When $\we <-1$, the region of stability always excludes the origin, which means that some non-minimal coupling \emph{needs} to be switched on in the presence of super-acceleration. 
Indeed, the upper-left panels of Figures~\ref{fig:stabeta20} and~\ref{fig:stabeta2} show that the Brans-Dicke theory allows for a stable phase of super-acceleration, a fact first noted in~\cite{Das:2005yj}. The more negative becomes the equation of state, and the greater the values of the non-minimal coupling $\alpha$ that one has to adopt in order 
to make the theory stable across all cosmic epochs. Super-acceleration seems particularly challenging for Brans-Dicke theories in the absence of other non-minimal couplings, because 
the lower limit  on acceptable $\alpha$ values  ($\alpha_{min}$) is strongly sensitive to the value of $\we$:  a small decrement of $\we$ 
translates into a large increase of $\alpha_{min}$.

 Also when the  high order operators  $\eta_3$ and $\eta_4$ are switched on (see panels {\it b),c)} and {\it d)} of Figure~\ref{fig:stabeta20}), a large value of $\alpha$ is generically requested in order  to make a theory stable. On the contrary, the parameter $\alpha$ can be  interpreted   as a small perturbative parameter  in the super-acceleration regime, if together with $\eta_3$ and $\eta_4$ we also switch on (and set to an extremely large value)  $\eta_2$. Indeed, by inspecting the last three panels of Figure
 \ref{fig:stabeta2},  we conclude that choosing a large value of $\eta_2$ is enough to enter the regime where $\we <-1$ with a relatively small effort. 
 
In the limiting case of large value of $\eta_2$, it is enough to switch on $\eta_3$ and/or $\eta_4$ individually to stabilize the super-acceleration phase. Consider, for instance, panel {\it b)} of Figure~\ref{fig:stabeta2}. Even by setting the Brans-Dicke coupling $\alpha$ to zero, stable theories can be found along the positive $\eta_3$-direction. In essence, this is the specific mechanism of super-acceleration studied in \cite{EFT1, Creminelli:2008wc} with the EFT formalism and in~\cite{deffa1} in terms of a specific model. The present work generalizes those
findings to a great extent by enlarging the dimensionality of theory space. For instance, we find that also the ``coordinate" $\eta_4$ can locate stable super-acceleration theories (Figure~\ref{fig:stabeta2}, panel {\it c)}), which might eventually be discriminated because of their different phenomenology (\emph{i.e.} because of their effects on the growth rate).

\section{Forecasts} \label{sec:5}
Suppose that the next generation of large scale surveys will measure 
the linear growth rate function of matter perturbation $f(t)$  finding that it is compatible with that of a $\Lambda$CDM model to a precision say of  $1\%$,  the nominal precision quoted by \cite{euclid2}.
How will  this observation constrain the space of alternative models for the propagation of gravity on large cosmic scales? Equivalently, which 
stable theories of modified gravity, if any,  will still be compatible  with  both background and perturbed sector data (on linear scales)?

\subsection{The growth index} \label{sec:5.1}
We address this issue by computing the growth index of linear density perturbations. 
Briefly, we assume that the quasi-static regime applies and  that the linear density perturbations of matter evolves as
\begin{align} \label{growthhhh}
\ddot \delta + 2 H \dot \delta - 4 \pi G_{\rm eff} \rho_m \delta = 0.
\end{align}

The EFT of dark energy makes characteristic predictions about the general form of the effective gravitational constant (see section 2.1.2 and eq.~\eqref{geff}). However, 
 the \emph{scale dependence} of $G_{\rm eff}$, which is encoded in the the infra-red corrective term $Y_{\rm IR}$ appearing both at the numerator and at the denominator of~\eqref{geff} and~\eqref{postn}, deserves
 a specific comment.  Our formalism allows a relatively compact explicit expression for $Y_{\rm IR}$, eq.~\eqref{ir}. Such a term becomes important at large distances---\emph{how large} depending crucially on the size of the non-minimal couplings.  As discussed at the beginning of Sec.~\ref{sec:3.3}, if the modified gravity mechanism plays a role in the cosmic acceleration, we generally expect the couplings to be of order Hubble to the appropriate power.
In this case, one can see by inspection that the infrared corrections become important only at wavelengths as large as Hubble itself, and thus fall outside the domain of validity of the quasi-static approximation. In other words, if the mechanism of modification of gravity is the one directly responsible for the acceleration of the Universe, the $Y_{\rm IR}$ term is generally irrelevant in the expressions~\eqref{geff} and~\eqref{postn} and the scale dependence effectively drops from such observables, leaving just an overall dependence on the time variable. Therefore, we will effectively ignore the $Y_{\rm IR}$ term in the present analysis.\footnote{The are specific limits in which the $k$-dependence is restored, but they represent a relatively narrow region of our parameter space. 
The case of $f(R)$ theories provides a notable example: if $\cb$ is strictly zero, $\mu \sim 10^{-3} H$ and all other couplings are null, then, by direct inspection of~\eqref{geff}, the $Y_{\rm IR}$ term starts becoming important well within the Hubble scale.} We also remark that  evolving the full dynamics of perturbations in linear regime,  without implementing any quasi-static approximation,  might eventually be required in view of the accuracy and
scale range of upcoming surveys \cite{HuRaFruSil}.

The differential equation~\eqref{growthhhh} translates into the following first order equation for the linear growth rate $f(t)=\frac{d \ln \delta}{d \ln a}$
\begin{equation}\label{f-equation}
3\bar{w} (1-x) x f'(x) + f(x)^2 + \left[2 + 3\bar{w} (1-x) x \frac{H'(x)}{H(x)}\right] = \frac{3x}{2} \frac{G_{\rm eff}(x)}{G_N} \, .
\end{equation}
Note that we are using $x$ as independent variable, the fractional non-relativistic matter energy density of the \emph{fiducial background model}, defined in eq.~\eqref{oe2}. A prime means differentiation with respect to it. On the RHS of the above equation, $G_{\rm eff}$ is time dependent, while the Newton constant $G_N$ is related to the present value of $M(t)$ through eq.~\eqref{refinement}.

It is a standard to parameterize $f$ by elevating to some power $\gamma$ the fractional matter density---which, consistently with our strategy so far, we take to be that of the fiducial background model,  $x$.  As shown in~\cite{SBM}, if $\gamma$ itself is expanded in powers of the logarithm of $x$, the approximation gets very precise for those models whose growth does not significantly deviate from that of $\Lambda$CDM. By taking just the first two terms of such an expansion we write 
\begin{equation}\label{eq:growthrate}
f(x)=x^{\gamma_0 + \gamma_1 \ln(x)}\, .
\end{equation}
This parameterization allows to compress the information contained in the function 
$f(t)$ in the two parameters $\gamma_0$ and $\gamma_1$, the so called growth indexes.  Once the function $H(t)$ and $G_{\rm eff}$ are specified in any given cosmological model, the  coefficients $\gamma_i$ can be straightforwardly  and fastly computed  using the prescriptions of \cite{SBM}.   
Any deviation of  the measured value of these coefficients $\gamma_i$  from the standard GR value is the smoking gun that gravitation may possess additional degrees of freedom.

\begin{figure}[t] \vspace{-1cm}
\centering
\includegraphics[trim = 2.75cm 6.5cm 2.75cm 6.5cm, scale=0.40]{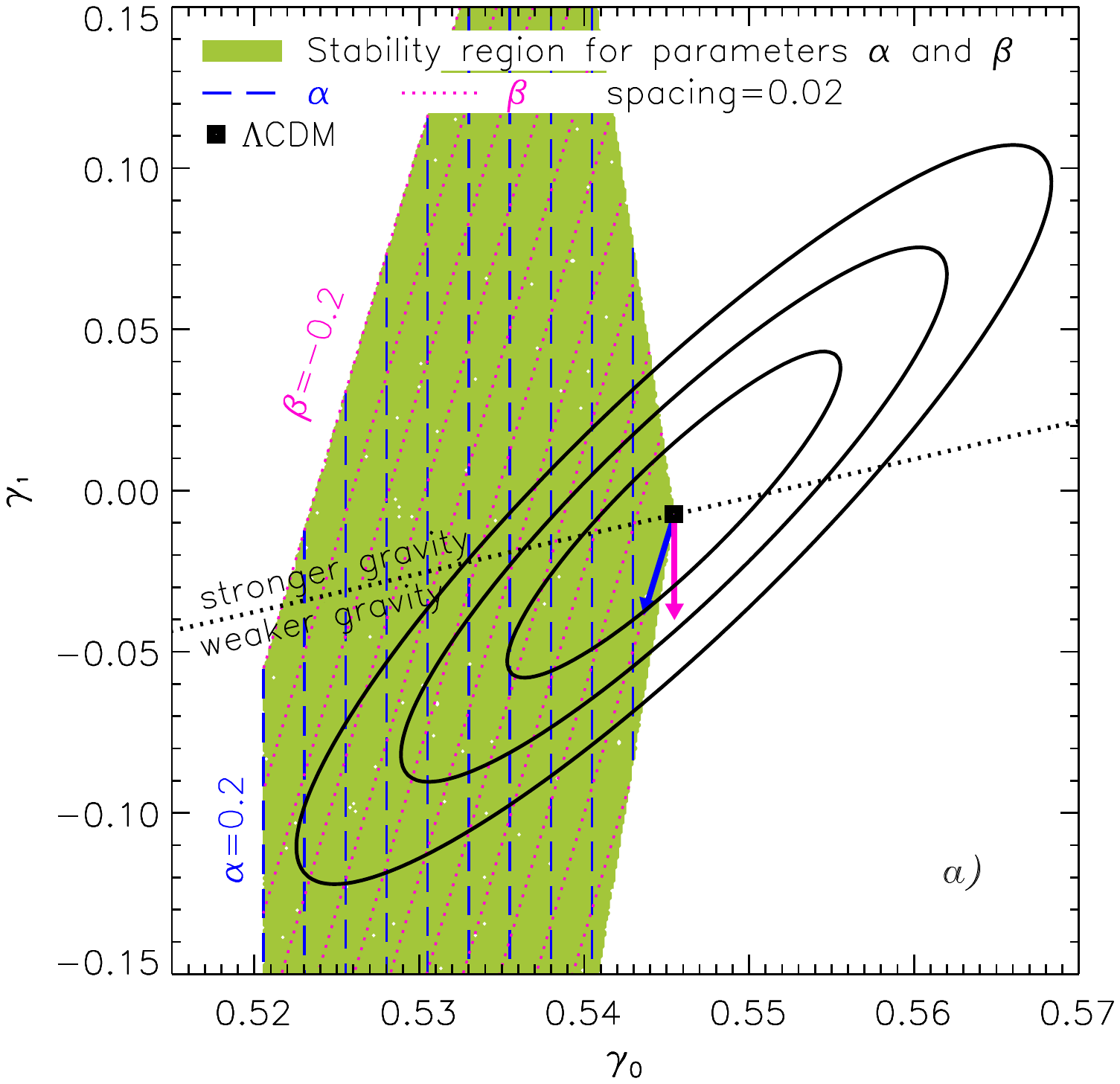}
\includegraphics[trim = 2.75cm 6.5cm 2.75cm 6.5cm, scale=0.40]{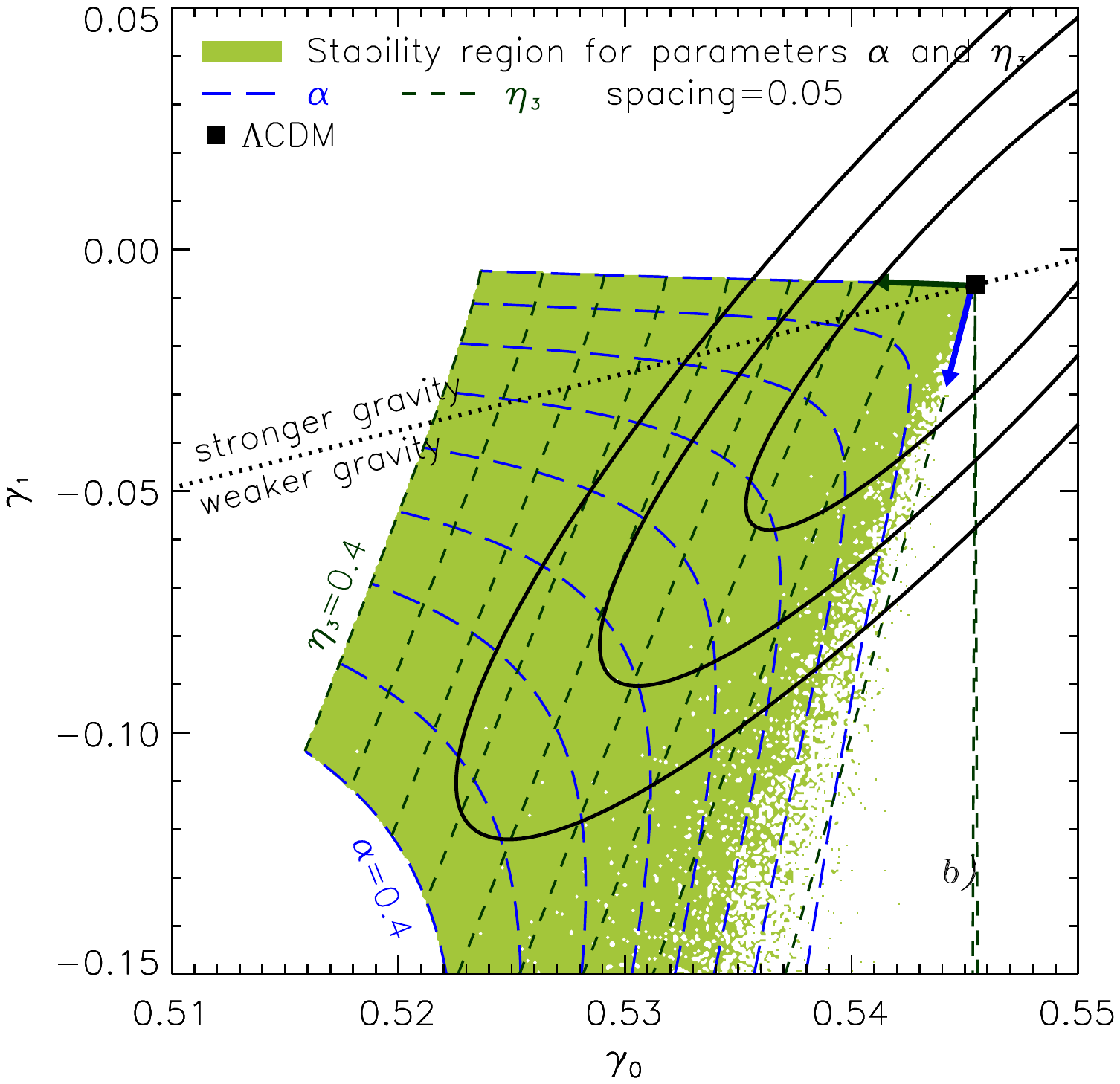} 
\includegraphics[trim = 2.75cm 6.5cm 2.75cm 6.5cm, scale=0.40]{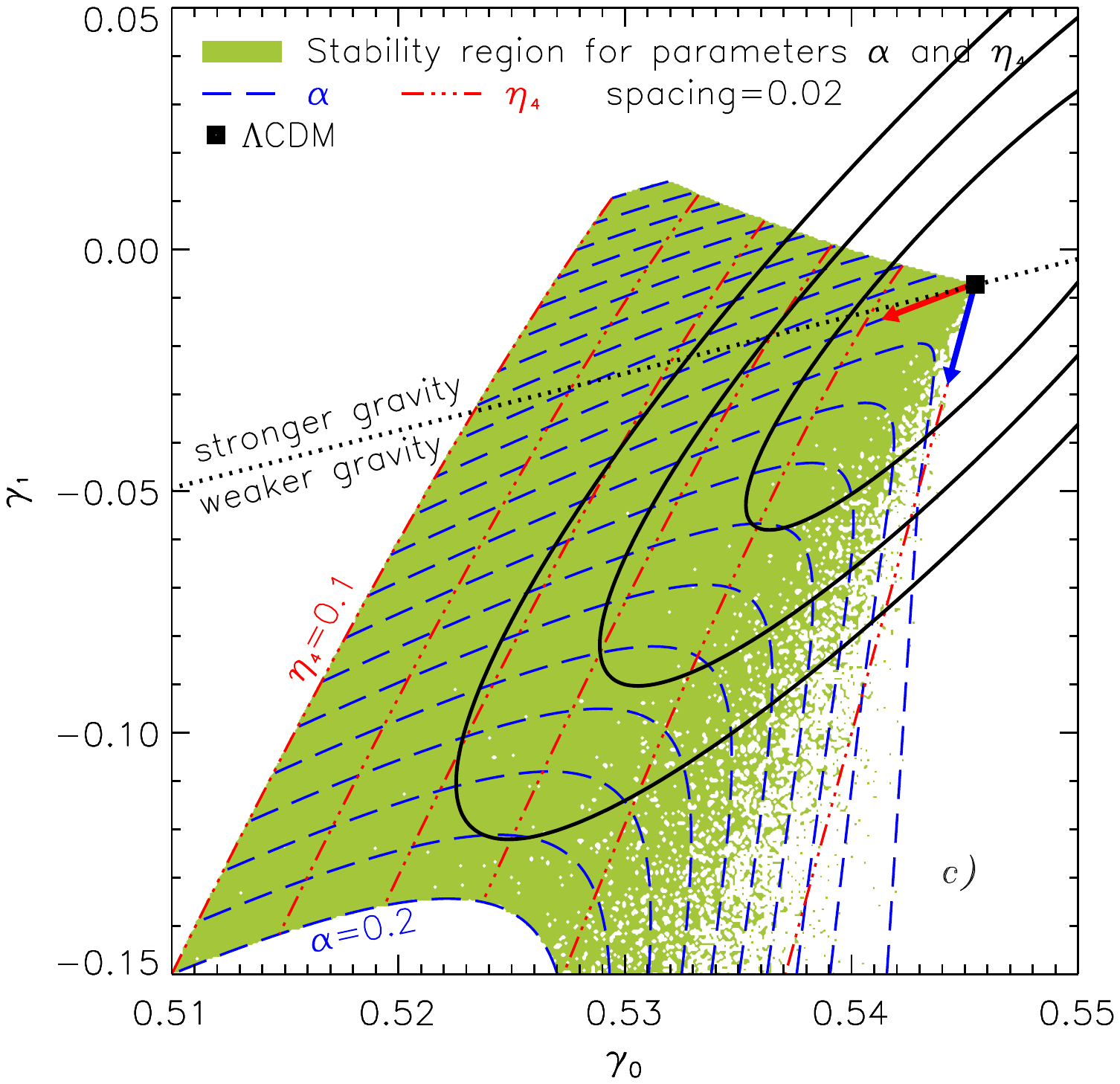}
\includegraphics[trim = 2.75cm 6.5cm 2.75cm 6.5cm, scale=0.40]{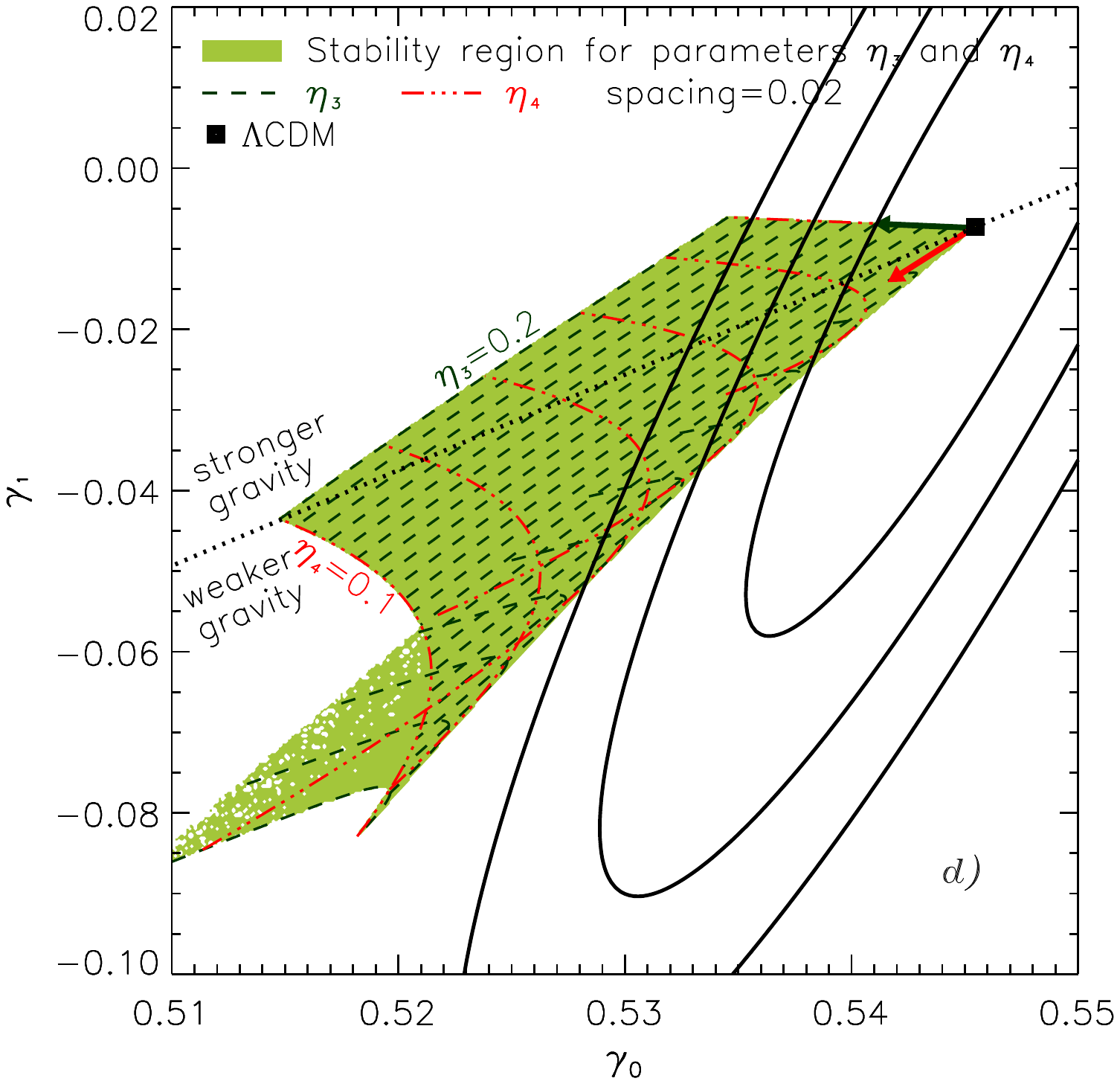} \vspace{0.5cm}
\caption{\label{fig:likg0g1}  
Expected $68\%$, $95\%$, and $99\%$ likelihood contours on the $\gamma_0$ and $\gamma_1$ parameters from an EUCLID-like survey. The black dotted line (``stronger/weaker gravity") divides the plane according to the present value of $f$, as calculated in~\eqref{eq:growthrate}. The $\Lambda$CDM model is shown by a black full square. Stabitility regions are shown in green and assume $\eta_2=0$.
{\it Upper left:} we also show the range of parameters $\alpha$ and $\beta$ which corresponds to  theories of gravity which are both stable and compatible with data.
{\it Upper right:} same as before but for theories specified in terms of the parameters $\alpha$ and $\eta_3$.
{\it Lower left:} same as before but for theories specified in terms of the parameters $\alpha$ and $\eta_4$.
{\it Lower right:} same as before but for theories specified in terms of the parameters $\eta_3$ and $\eta_4$.
}  
\end{figure}

We simulate future  Euclid measurements  of the growth rate $f(t)$ as explained in \cite{SBM}.  Essentially,  we adopt the figure of merit 
quoted in  \cite{euclid2} and  assume as fiducial a $\we$CDM model with parameters $x_0=0.3$ and $\we=-1$.
We then perform a maximum likelihood analysis of data with  the  model given in  eq. \eqref{eq:growthrate}.
\begin{figure}[t] \vspace{-1cm}
\centering
\includegraphics[trim = 3cm 5.5cm 3cm 3cm, scale=0.40]{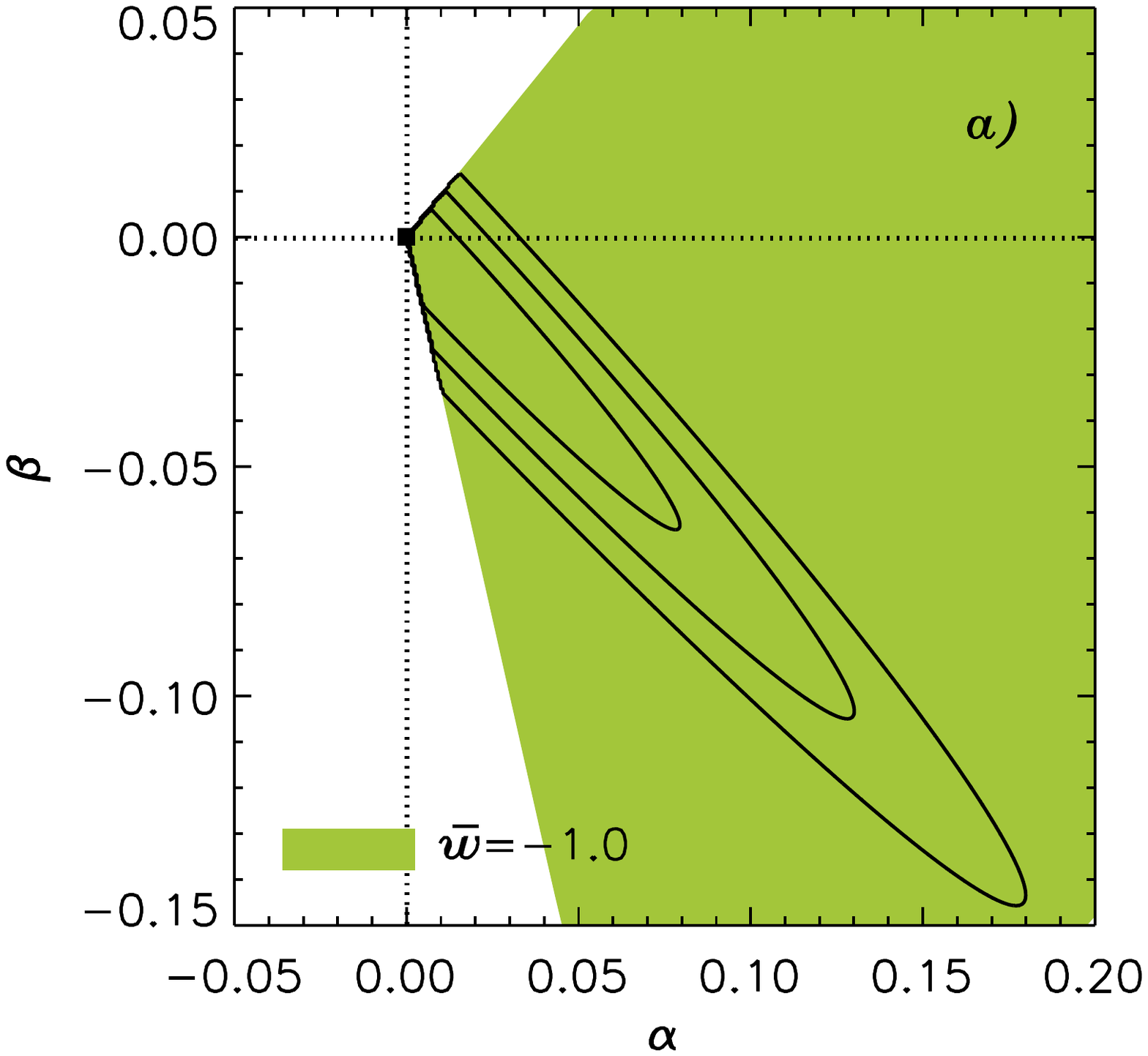}
\includegraphics[trim = 3cm 5.5cm 3cm 3cm, scale=0.40]{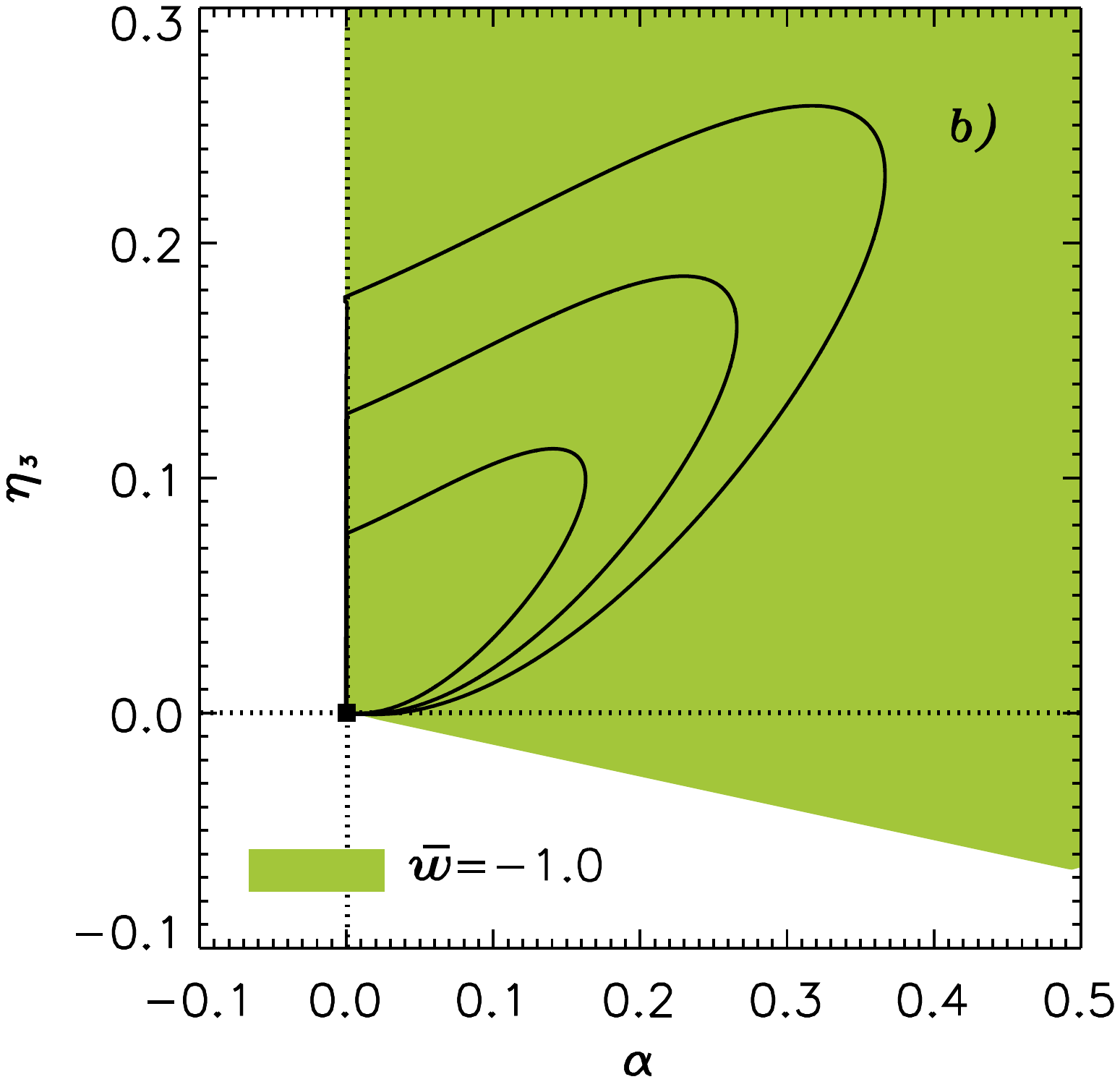} 
\includegraphics[trim = 3cm 3cm 3cm 5.5cm, scale=0.40]{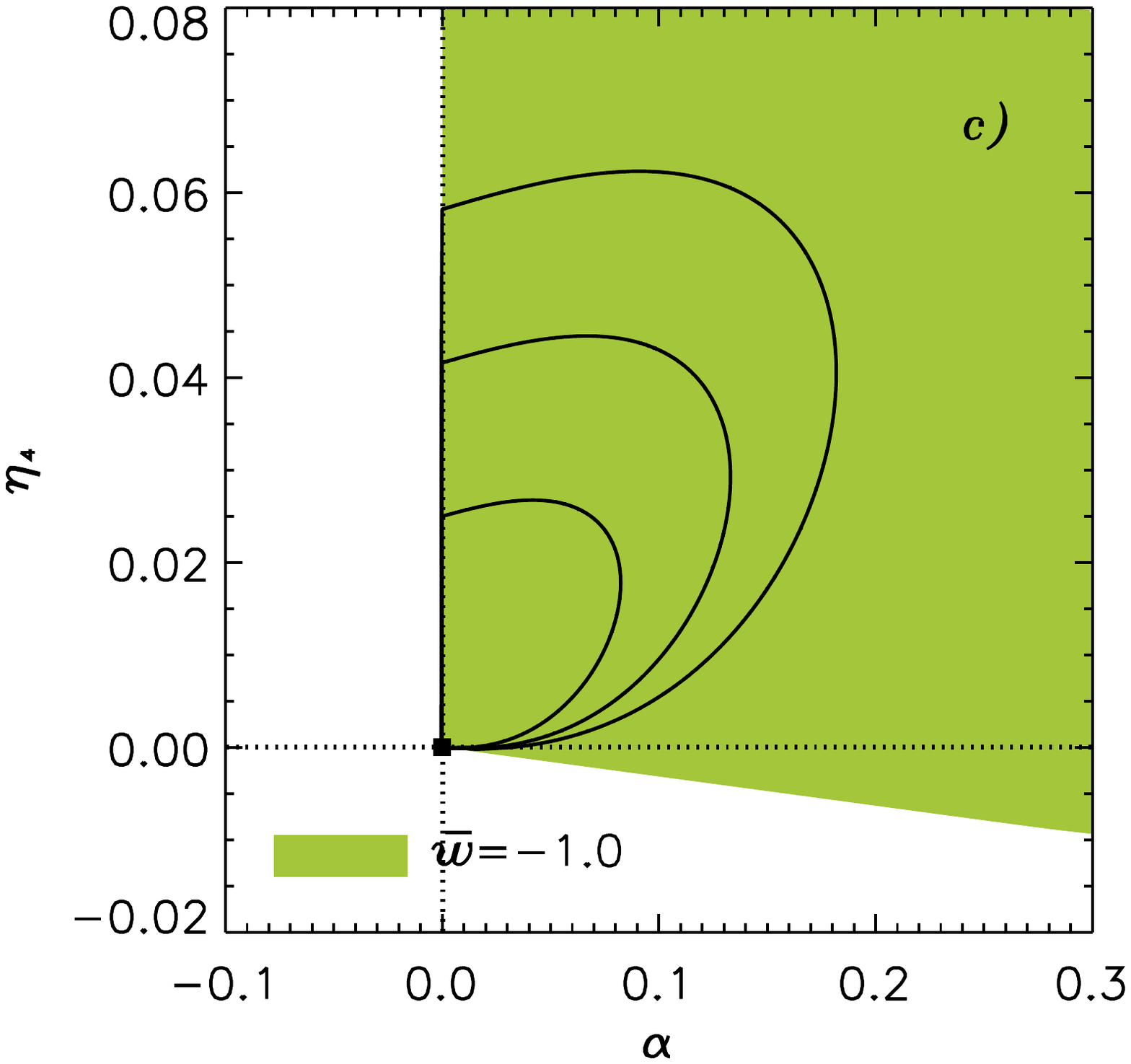}
\includegraphics[trim = 3cm 3cm 3cm 5.5cm, scale=0.40]{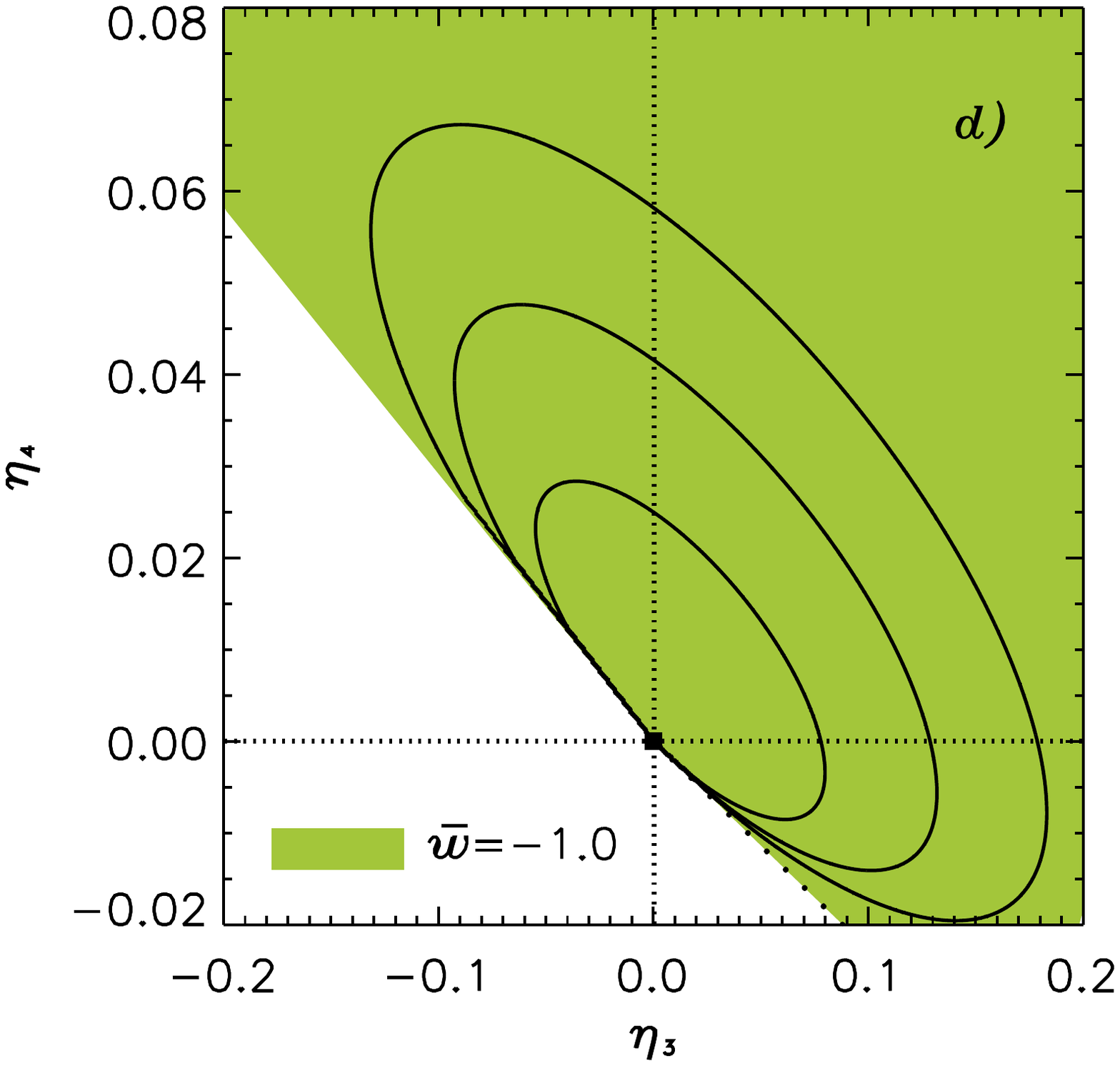} \vspace{-1cm}\caption{\label{fig:likabE}  In each panel, the shaded area represents the region of stability for EFT theories predicting an expansion history given by the flat $\we$CDM model with parameters 
$x=0.3$ and $\we=-1$. The contours represent  the projection of the $68\%$, $95\%$, and $99\%$ likelihood levels for the $\gamma_0$ and $\gamma_1$ parameters
(shown in Figure \ref{fig:likg0g1})  onto the plane $\alpha- \beta$  ({\it Upper left})  $\alpha-\eta_3$  ({\it Upper right:}), $\alpha-\eta_4$ ({\it Lower left}) and $\eta_3-\eta_4$  ({\it Lower right}).
}  
\end{figure}
The resulting  likelihood contours for the growth indexes $\gamma_0$ and $\gamma_1$  are shown in Figure \ref{fig:likg0g1}.
In this same figure we also show the EFT  predictions for the amplitude of  $\gamma_0$ and $\gamma_1$ which are compatible  with both  expected data
and  requirements of theoretical stability. 
We stress that, given our background-perturbation separation strategy,  any stable EFT model (parameterized in terms of  $\alpha, \beta, \eta_3$ and $\eta_4$) 
lying within the likelihood contours of growth rate data, is also a model which identically reproduces the expansion rate of the fiducial  $\we$CDM model. 
Figure \ref{fig:likg0g1}  deserves a few comments. 

\begin{itemize}
\item It appears that  Brans-Dicke like models offer a maximum coverage of the likelihood surface, with parameters $\alpha$ and $\beta$ varying roughly
in the range $[0,0.2]$  and $[0,0.15]$ respectively. On the contrary, the space of  viable theories parameterized by $\eta_3$ and $\eta_4$ is much more 
constrained, as can be seen in panel {\it d)}.

\item By opportunely choosing the EFT  parameters in each panel,  the growth rate $f$ can be made larger or smaller than the growth rate in the fiducial model. The black dotted line,  drawn 
as explained in \cite{SBM},  divides the $\gamma_0-\gamma_1$ plane according to whether local gravity,  i.e. the mechanism responsible for the  linear growth of structures,  is stronger or weaker than in  the fiducial  $\Lambda$CDM  model. For example, in Brans-Dicke like models, in which an increase of $\alpha(/\beta)$ produces a decrease(/increase) of $\gamma_0(/\gamma_1)$, theoretically  stable and observationally viable theories may generate a present  day growth  rate $f$ that is  $12\%$ smaller and $5\%$ higher  than that of the fiducial model.
Note, however,  that in the vast majority of cases, stable theories predicting  a background expansion rate identical to that of the $\Lambda$CDM model, also predict  growth suppression mechanisms
that are more efficient than the cosmological constant. In other terms,  it  is much more likely that a theory which is  undistinguishable from the $\Lambda$CDM model 
as far as background observables are concerned, predict a different, \emph{lower}, growth rate of cosmic structures if it is to be in agreement with observations in the perturbed sector.

\item Our parameterization suggests that stable models of modified gravity  cannot extend in regions where  the zeroth order growth index $\gamma_0$  is more positive than the $\Lambda$CDM 
value. In other terms, the stability condition sets a limit on the maximum growth index $\gamma_0$ a theory can have. 
In principle, this might be an artefact due to the specific parameterization adopted. In Sec.~\ref{sec:5.2} we show that this is indeed a universal feature of viable modified gravity models.

\end{itemize}

Overall,  Figure \ref{fig:likg0g1}  illustrates a central result of this paper,  {\it i.e.}  the space spanned by stable gravitational theories that are not statistically rejected by data is actually much smaller than 
that enclosed by the empirical  likelihood contours. By this,  we demonstrate the importance of analyzing data with a general  EFT-like formalism: the figure of merit in the $\gamma_0-\gamma_1$ plane
is naturally boosted, not only by increased observational capabilities, but also by enhanced theoretical understanding. 
Given that we will show in Section 2 that the region for $\gamma_0>\gamma_{0,\Lambda CDM}$ cannot be described by any stable theory with the same background expansion of the $\Lambda$CDM
model, the gross features of the confidence region which is theoretically unbiased can be safely considered as independent from the adopted  parameterization scheme.
The finer details of the bounds imposed by theory, however,  might be parameter-dependent.   The extent to which this is affecting our conclusions  will be  explored in a forthcoming  paper.  

While Figure \ref{fig:likg0g1} is more observer friendly, in the sense that it projects theoretical results about the amplitude of the EFT parameters directly in the
plane of cosmological observables (i.e. $\gamma_0$ and $\gamma_1$), Fig~\ref{fig:likabE} is, in the same spirit, more theorist friendly, since it projects the 
growth index  likelihood contours directly in the phase space of modified gravity theories. From this plot one can straightforwardly deduce the 
set of modified gravity theories which are in  agreement with results of the simulated experiment. For example,  while no positive $\beta$ values can be paired to a positive 
$\alpha$ parameter, the contrary happens to  $\eta_3$ and $\eta_4$, which, once paired with $\alpha$,   fit data  only if their value is negative. 

\begin{figure}[t] \vspace{-1cm}
\centering
\includegraphics[trim = 3cm 5.5cm 3cm 3cm, scale=0.40]{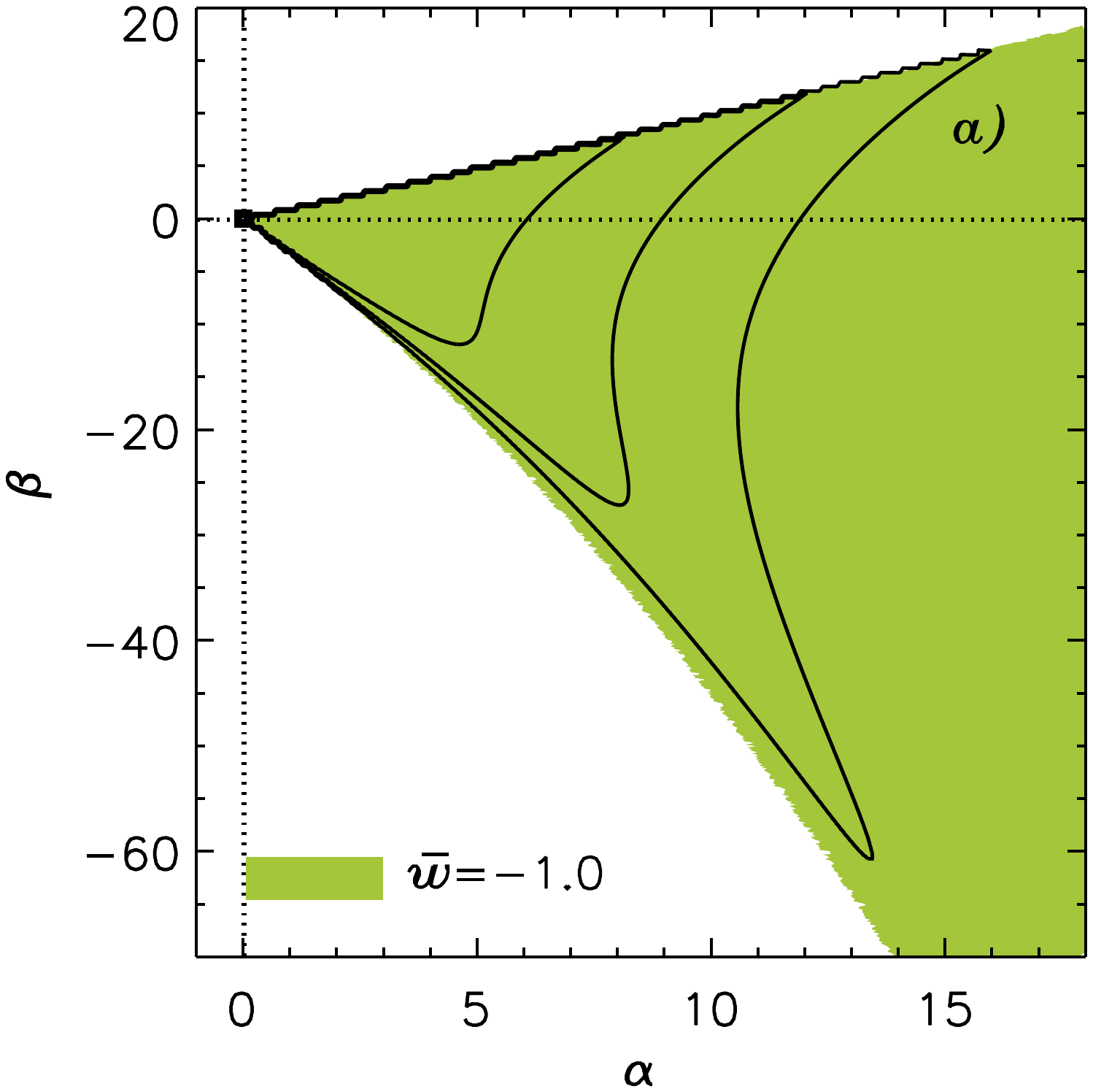}
\includegraphics[trim = 3cm 5.5cm 3cm 3cm, scale=0.40]{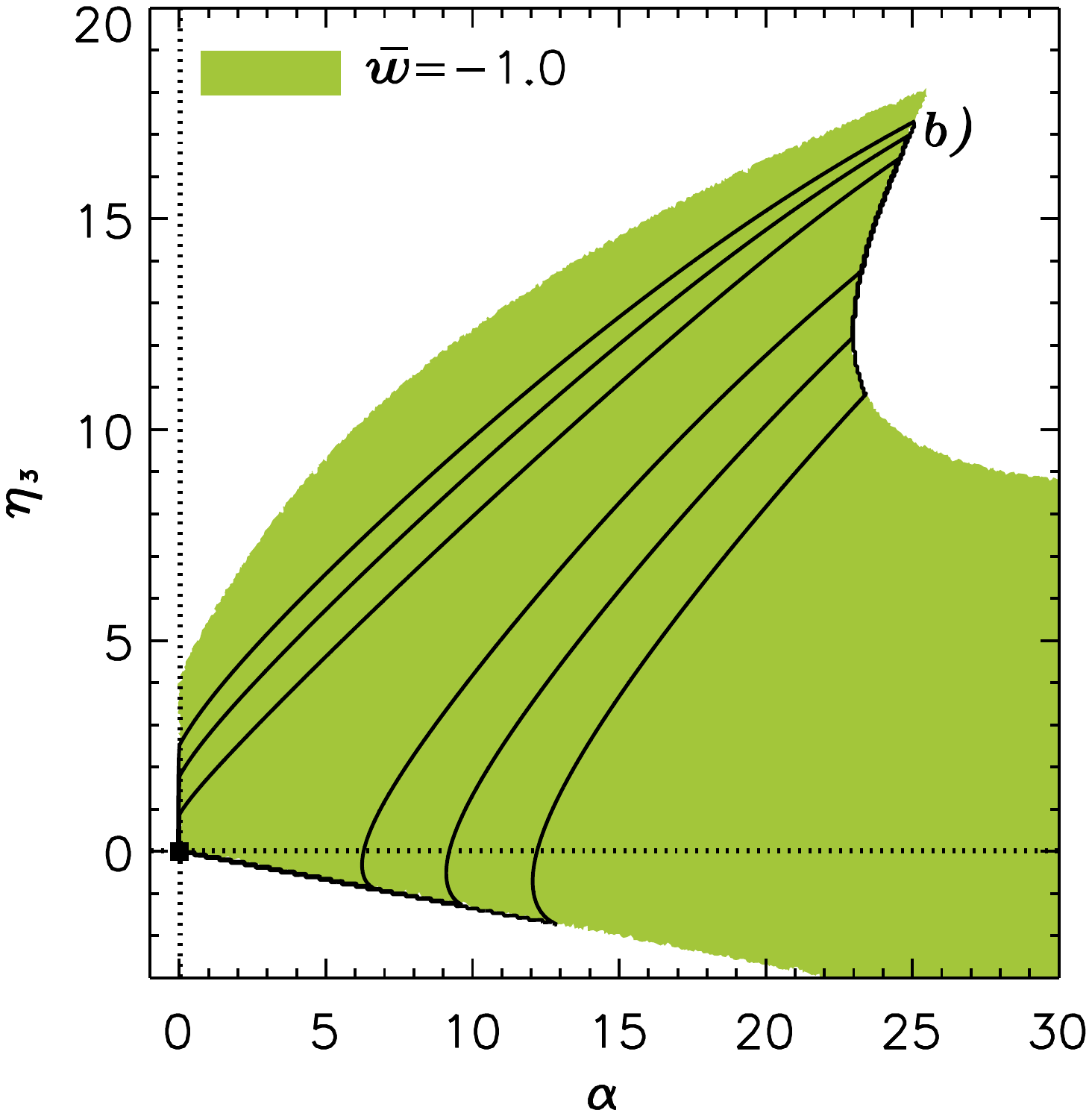} 
\includegraphics[trim = 3cm 3cm 3cm 5.5cm, scale=0.40]{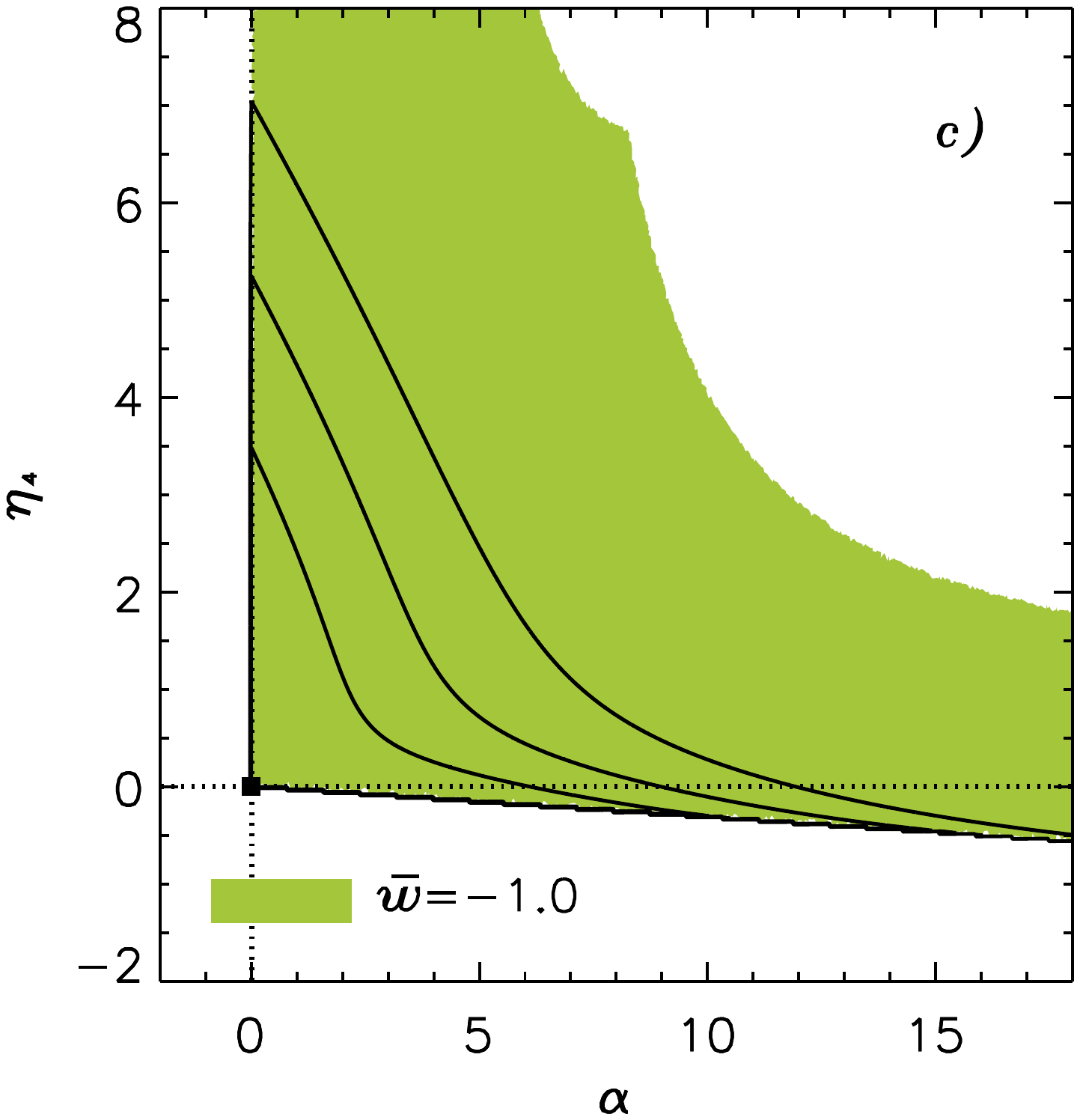}
\includegraphics[trim = 3cm 3cm 3cm 5.5cm, scale=0.40]{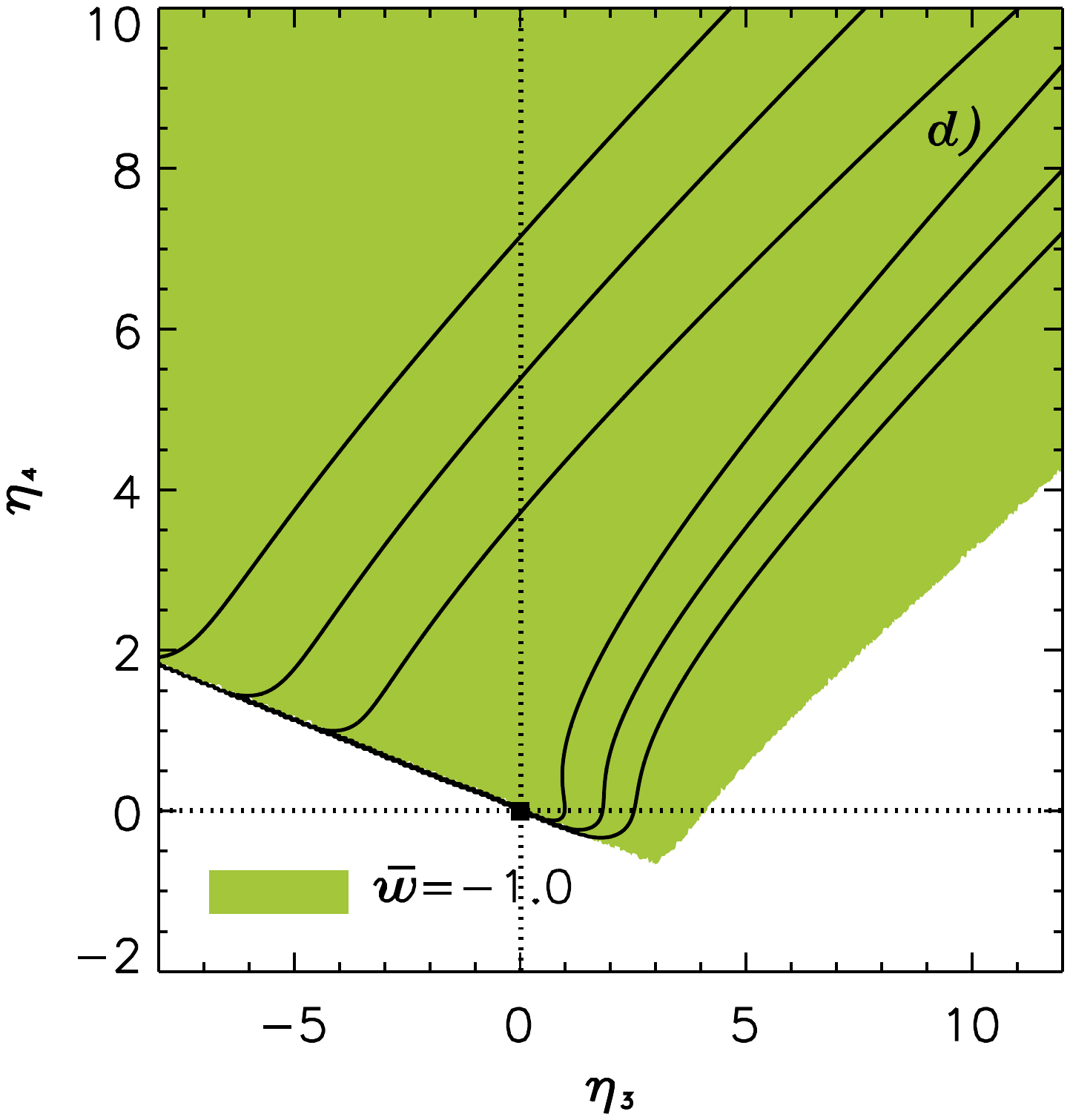} \vspace{-1cm}
\caption{\label{fig:likab}  
  In each panel, the shaded area represents the stability region of  EFT theories predicting an expansion rate history which is indistinguishable from that expected in a $\we$CDM model with parameters 
$x_0=0.3$ and $\we=-1$.  The contours represent  the projection of the $68\%$, $95\%$, and $99\%$ likelihood levels for the $\gamma_0$ and $\gamma_1$ parameters
onto the plane $\alpha- \beta$  ({\it Upper left})  $\alpha-\eta_3$  ({\it Upper right:}), $\alpha-\eta_4$ ({\it Lower left}) and $\eta_3-\eta_4$  ({\it Lower right}). They are  obtained 
from available measurements of the growth rate history    \cite{Guzzo2008,SonPer09, DavNusMas11, BlaGlaDav11, ReiSamWhi12, SamPerRac12, BeuBlaCol12, TurHudFel12} as detailed in 
\cite{SBM}.}
\end{figure}

Finally, in Figure \ref{fig:likab}  we  show the EFT parameters that are compatible with current data about the growth rate of structures. 
To this purpose we analyze the measurements of   \cite{Guzzo2008,SonPer09, DavNusMas11, BlaGlaDav11, ReiSamWhi12, SamPerRac12, BeuBlaCol12, TurHudFel12}  
using the prescriptions detailed in  \cite{SBM}. Statistical degeneracy affects the parameters $\alpha-\eta_3$ and $\eta_3-\eta_4$. While in the former case the degeneracy is 
resolved by imposing stability conditions, in the latter it can be overcome only by increasing quality and quantity of astronomical data, as shown in Figure \ref{fig:likabE}.

\subsection{$\Lambda$CDM vs. modified gravity: comparing the growth rates} \label{sec:5.2}

Since gravity---even when modified---is an attractive force on small scales, one could naively expect  that the growth of structures is always enhanced 
when a non-minimal coupling is switched on in the theory.  However, the complexity of the system does not allow us to draw such a  universal statement. 
Indeed, as outlined in the comments to Figure~\ref{fig:likg0g1}, the vast majority of stable theories seems to produce an overall growth of structures today, $f(t_0)$, which is \emph{smaller}  than that produced by $\Lambda$CDM.  

Still, there is a sense in which ``larger couplings" imply ``more growth".  Of the growth indexes introduced in the last section, $\gamma_0$ is the one relevant at very early times, \emph{i.e.} at the onset of dark energy domination. While the total growth function today, $f(t_0)$, is the result of the entire time evolution, $\gamma_0$ is only sensitive to the ``initial kick" given by  the new component---the accelerating mechanism at the time when it starts to be effective. Our plots of Figure~\ref{fig:likg0g1} suggest that $\Lambda$CDM always correspond to the highest allowed value of $\gamma_0$---and therefore to the lowest ``initial tendency" to structure growth---among all the models with the same effective equation of state $\we = -1$. 
 Indeed, the areas of stability are always on the left of the point representing $\Lambda$CDM.  This, however, could be an artifact of our specific parameterization as well.

Here we show that, irrespectively of the adopted parameterization scheme,  the $\Lambda$CDM model always maximizes $\gamma_0$. This  follows straightforwardly from one of the most noticeable properties
of the proposed formalism, the possibility of a direct comparison of  the growth rates of disparate dark energy theories \emph{which share the same effective equation of state} $\we$. 
Let us first consider the Brans-Dicke sector of the theory. Models with a given effective equation of state $\we=const.$ have a specific relation between $w(x)$ and $\eta(x)$, which is given in eq.~\eqref{etaaa}. 
Here, at variance with what we did  in Sec.~\ref{sec:4} we choose to define the Brans-Dicke model by means of $w(x)$, and to calculate $\eta(x)$ accordingly.  Specifically, we  define 
\begin{equation}
\ell(x) \ = \ w(x) \,- \, \we \, .
\end{equation}
and note that the parameter $\gamma_0$ depends only on the value of $\ell$   at $x = 1$, \emph{i.e.} $\ell_1 \equiv \ell (1)$. For $\we =-1$ we find
\begin{equation}
\gamma_0\ = \ \frac{6}{11}\, \frac{1 - 2 \ell_1}{1-\ell_1} \ = \ \frac{6}{11} \, -\, \frac{6 \ell_1}{11} \, +\, {\cal O}(\ell_1^2)\, .
\end{equation}
We can see that $\gamma_0$ is a decreasing function of $\ell_1$, which  reaches the $\Lambda$CDM value $\frac{6}{11}$ at $\ell_1 = 0$. 

On the other hand, the no-ghost and gradient-stabilities conditions~\eqref{conditions} coincide in the Brans-Dicke case, and can be expanded in powers of $(1-x)$ at early times. The leading term is linear in $(1-x)$. By requiring its positivity we obtain the stability condition
\begin{equation} \label{newnoghost}
\we \, \frac{2 + 2\we + 7 \ell_1 + 6 \we \ell_1}{\we + \ell_1}\ \geq \ 0\, .
\end{equation}
For $\we = -1$ the above condition reduces to $0 \leq  \ell_1 < 1$. So, if $w$ gets not too far from $\we$ at $x = 1$, the stability condition does  imply $\gamma_0 \leq \frac{6}{11}$.

If also the other coupling functions $\eta_3(x)$ and $\eta_4(x)$  are switched on, the expression for $\gamma_0$ still depends only on the value of such couplings at $ x = 1$,  that is, on the amplitudes $\eta_3(1)$ and $\eta_4(1)$ (see Appendix A). By expanding   the stability conditions  around $x = 1$ we find that the no-ghost condition~\eqref{noghost} is identical to~\eqref{newnoghost}. On the other hand,  the gradient-stability condition~\eqref{gradient} becomes  more involved but still depends only on the value $\eta_3(1)$ and $\eta_4(1)$ (and not \emph{e.g.} on their derivatives):  the stability problem is still an algebraic inequality among a finite number of parameters.  For $\we = -1$ we have verified numerically that the stability conditions, also in the most general case, imply
\begin{equation}
\gamma_0 \ \leq  \ \frac{6}{11}\, .
\end{equation}
In summary, for a fixed effective equation of state parameter $\we = -1$,  the $\Lambda$CDM model maximizes the allowed  values of the leading order growth index $\gamma_0$. 
We should notice that while smaller $\gamma_0$ means stronger gravity ``at the beginning", \emph{i.e.} during matter domination, the region with a smaller $\gamma_0$ does not imply a larger value of the total growth rate now, $f(t_0)$. This is visible from the stronger/weaker gravity line drawn in Figure \ref{fig:likg0g1}. Indeed, as a tendency, we find that the presence of non-minimal couplings strongly favors a weaker overall growth, despite the initial positive kick given by a lower value of $\gamma_0$. 

\section{Conclusions}

The effective field theory of dark energy provides a framework for parameterizing  
possible departures from the standard gravitational paradigm on large scales and, at the same time,  interpret eventual non-null detections  in terms of fundamental gravitational proposals. 
Its most appealing features are twofold: {\it a)} the effectiveness with which a general class of  gravitational models obtained by adding a single degree of freedom to general relativity 
can be unified and  their predictions systematically compared to data, and  {\it b)} the straightforward identification and classification of the   operators controlling the evolution of the background and of the (linearly) perturbed sector of the universe.

In this paper we have shown that  the six universal couplings entering the EFT Lagrangian ($M^2, \lambda,  \cb, \mu_2,  \mu_3$ and $\epsilon_4$, eq. \eqref{example}) can be re-expressed 
in terms of five dimensionless functions ($\we$ the effective dark energy equation of state of a Friedmann model of the universe, 
the non-minimal ``Brans-Dicke" coupling $\eta$, and high order couplings $\eta_2, \eta_3$ and $\eta_4$, see Table~\ref{symbols}) such that 
operators responsible for the expansion  and growth histories are distinct and independent.  Indeed, while the effective equation of state parameter $\we$ depends only on the expansion rate of the universe, the four remaining functions are only active in the  perturbation sector. Among them, only three ($\eta, \eta_3$ and $\eta_4$)  have a direct influence on the growth rate of cosmological structures at the linear level, and may be responsible
for non-standard structure formation processes.

As a convenient---but by no means univocal---choice, we propose to parameterize these functions in terms of a set of six coefficients (see Table~\ref{symbols2}, right column). We show that  this parameterization scheme is general enough to cover most interesting deviations from standard gravity, and flexible enough to 
encompass most theories without  pathologies, that is, free from ghost or gradient instabilities. We then use these coefficients  as ``coordinates"  to locate potentially viable gravitational theories
and test for a non-minimal coupling of the dark sector to gravity (in the Jordan frame). We show that the  volume spanned by  non-pathological  theories is progressively reduced as 
the dark energy equation of state parameter decreases towards negative values.  Such a ``theory-space behavior",  although somewhat expected, has never been quantified before to our knowledge, but can be easily tracked in our formalism because $\we$ and the non-minimal couplings $\eta$-$\eta_i$ are treated as independent quantities. 

Specifically, no minimally coupled scalar field  can generate  a super-accelerated expansion  (i.e. $\we <-1$).  However, in the presence of a sufficiently large $\eta_2$, even a small value of the $\eta_4$ parameter with all other couplings set to zero  is able to stabilize theories with a strongly negative $\we$ (Figure~\ref{fig:stabeta2}, bottom-right panel). The parameter $\eta_4$ is typical~\cite{GLPV} of the higher-order Galilean Lagrangians (the ``${\cal L}_4$" and ``${\cal L}_5$" terms)~\cite{NRT} and their generalized versions~\cite{horndeski,Deffayet:2009mn}. We also expect it in Gauss-Bonnet $f(G)$-type theories~\cite{Nojiri:2005jg}. 

Besides being instrumental in searching  for explicitly  covariant models that comply with stability constraints, the formalism developed in this papers also serve as a guide in  
interpreting  empirical results about the amplitude of relevant cosmological observables.  Future  surveys of the LSS  are expected  to constrain, with unprecedented precision,  both geometrical (smooth)
and dynamical (perturbed) observables of the cosmological model.  In particular the Euclid survey is expected to test  the large scale limit of Einstein
gravity by measuring the growth index $\gamma$  to a 1-sigma precision of $< 0.02$.
We  rely on  this predicted figure of merit to forecast how the parameter space of viable gravitational alternatives will shrink under data pressure.
We show that likelihood contours for the parameters  $\gamma_0-\gamma_1$  have a purely formal, phenomenological  nature.
Since  the growth index is a model dependent quantity, the statistical  limits on its amplitude,  if not properly interpreted using a general gravitational  formalism   such as the EFT, 
overestimate the true range of theoretically allowed values.   Indeed,  we  have demonstrated  that  only a fraction of the statistically allowed region is also physically viable, 
i.e.  it is spanned by stable theories of gravity. In particular we have found (Sec.~\ref{sec:5.2}) that once the cosmic expansion rate is fixed to $\we = -1$,
no viable theory can generate a leading order growth index $\gamma_0$  which is larger than that of $\Lambda$CDM. 

Most of  our conclusions rest on the specific parameterization that we have adopted in order to turn the phenomenological exploration of an additional gravitational degree of freedom 
into a tractable problem.  A  point  that  needs to be further investigated is thus  the degree of generality guaranteed by such a parameterization scheme. 
Despite the  proofs that, irrespectively of the adopted parameterization,  $\gamma_0$  for $\we = -1$  is always lower than that predicted in a  
$\Lambda$CDM model, one is still left with the issue of investigating  whether our specific parameterization offers the maximal possible coverage of the empirical  likelihood 
in the plane  $\gamma_0$-$\gamma_1$. 
Another  line of investigation concerns the application of the EFT formalism to interpret perturbation observables other than the growth rate.  For example, it would be interesting to 
work out which  theoretical constraints fundamental physics imposes on the amplitude of the gravitational slip parameters.

\section*{Acknowledgments}
\allowdisplaybreaks[1]
We acknowledge useful discussions with Luca Amendola, Julien Bel, Paolo Creminelli, Noemi Fru\-sci\-ante, Jerome Gleyzes, Luigi Guzzo, Justin Khoury, David Langlois, Marco Raveri, Alessandra Silvestri, Enrico Trincherini and Filippo Vernizzi. FP acknowledges the financial support of the UnivEarthS Labex program at Sorbonne Paris Cit\'e (ANR-10-LABX-0023 and ANR-11-IDEX-0005-02).
CM is grateful for support from specific project funding of the {\it Institut Universitaire de France} and of the Labex OCEVU.
\appendix
\section{Expressions of $\gamma_0$ and $\gamma_1$}
\label{sec:annexe} 
We give the expressions of $\gamma_0$ and $\gamma_1$ for the Brans-Dicke sector of the theory (couplings $\alpha$ and $\beta$):
\begin{align}
\gamma_0 = & \; \frac{3(1-\we)}{5-6\we} - \frac{\alpha(1-\xo)\big[6\we (1+\we)-\alpha(1-\xo)(2-\we(5+9\we)) \big]}{\we\big[6\we(5-\we +6\we^2)-\alpha(1-\xo)(10-37\we +36\we^3) \big]} \\[2mm]
\gamma_1 = & - \Big[ 8\alpha^4(1-\xo)^4 + 4\we\alpha^2(1-\xo)^2\big(\alpha(51+99\xo)+14\alpha^2(1-\xo)^2+150\beta(1-\xo)\big) \nonumber \\ 
&  + 972\we^8(1+\alpha(1-\xo))^2 +324\we^7\Big(1+12\alpha^3(1-\xo)^2(1+\xo)+24\beta(1-\xo)  \nonumber \\
& + 6\alpha^2(1-\xo)\big(9+\xo(7-8\beta)+4\beta(1+\xo^2)\big) + \alpha\big(13+\xo(11-192\beta)+96\beta(1+\xo^2) \big) \Big) \nonumber \\
& + 4\we^2\alpha(1-\xo)\Big(\alpha^2(603-96\xo -507\xo^2)+199\alpha^3(1-\xo)^3 \nonumber \\
& -900\beta(1-\xo) -6\alpha\big(57+\xo(93-370\beta)+185\beta(1+\xo^2) \big) \Big) \nonumber \\
& - 9\we^5\Big(36 + 324\alpha^4(1-\xo)^4 + 1416\beta(1-\xo)  + 24\alpha\big(25+\xo(34-408\beta)+204\beta(1+\xo^2)\big)\nonumber \\ 
& +6\alpha^3(1-\xo)^2(385+183\xo) +(1-\xo)\alpha^2\big(2615+\xo(2281-6816\beta)+3408\beta(1+\xo^2)\big)\Big)  \nonumber \\
& +27\we^6\Big(24\alpha^3(1-\xo)^2(3+7\xo) -72\alpha^4(1-\xo)^4 -32\alpha\big(2+9\beta(1+\xo^2)-\xo(5+18\beta)\big) \nonumber \\
&  -12(5-8\beta(1-\xo)) -3\alpha^2(1-\xo)\big(85-80\beta(1+\xo^2)+\xo(11+160\beta) \big) \Big) \nonumber \\
& +9\we^4\Big(\alpha^3(1-\xo)^2(1193+507\xo)+353\alpha^4(1-\xo)^4 +24(3-10\beta(1-\xo)) \nonumber \\  
& +4\alpha^2(1-\xo)\big(221+425\beta(1+\xo^2)+25\xo(5-34\beta)\big) + 8\alpha\big(6+173\beta(1+\xo^2)-2\xo(18+173\beta) \big)  \Big) \nonumber \\
& +3\we^3\Big(\alpha^3(1-\xo)^2(1507+331\xo) +332\alpha^4(1-\xo)^4 +36\alpha\big(17+\xo(33-260\beta)+130\beta(1+\xo^2)\big) \nonumber \\ 
& +1800\beta(1-\xo) +2\alpha^2(1-\xo)\big(1104+919\beta(1+\xo^2)+2\xo(618-919\beta) \big) \Big) \Big] \nonumber \\
& / \Big[ 6\we^2(5-12\we)(5-6\we)^2\big(2\alpha(1-\xo) - \we(6+5\alpha(1-\xo)) -6\we ^2(1+\alpha(1-\xo))\big)^2 \Big]
\end{align}
When all couplings are switched on, we find that the expression of $\gamma_1$ is very involved. That of $\gamma_0$, on the other hand, can be given directly in terms of $w$, $\eta_3$ and $\eta_4$ calculated at $x=1$:
\begin{align}
\gamma_0 = & \Big[3\Big(9\we^4(3-2w_{1})-2\we w_{1}\big((1-4w_{1})\eta_{31}+(11+10w_{1})\eta_{41}\big) +w_{1}^2(\eta_{31}^2-16\eta_{41}^2) \nonumber \\
 & +3\we^3\big(5+6w_{1}^2-w_{1}(17-24\eta_{41})\big)  - 3\we^2\big(2+w_{1}(5-10\eta_{41})-2w_{1}^2(5-\eta_{31}-14\eta_{41}+6\eta_{41}^2)\big)  \Big)\Big]  \nonumber \\
 & / \Big[ w_{1}(5-6\we)\big(18\we^3 + 3\we^2(5-6w_{1})-2w_{1}(\eta_{31}+5\eta_{41})- 3\we(2+w_{1}(7-2\eta_{31}-4\eta_{41})) \big)\Big]
\end{align}
where $w_1=w(1)$, $\eta_{31}=\eta_3(1)$ and $\eta_{41}=\eta_4(1)$.

\end{document}